\documentclass[%
    aip,
    jcp,
    amsmath,
    amssymb,
    twocolumn,
    superscriptaddress,
    reprint]{revtex4-1}

\usepackage{graphicx}%
\usepackage{dcolumn}
\usepackage{amsmath,mathtools}   
\usepackage[usenames,dvipsnames]{pstricks}
\usepackage{pst-grad} 
\usepackage{pst-plot} 
\usepackage[small,compact]{titlesec} 
\usepackage{bm}
\usepackage{color}
\usepackage[low-sup]{subdepth}

\begin{document}

\title{Electrode Reactions in Slowly Relaxing Media }

\author{Dmitry V.\ Matyushov }
\affiliation{Department of Physics and School of Molecular Sciences, Arizona State University,  PO Box 871504, Tempe, AZ 85287-1504 }
\email{dmitrym@asu.edu}

\author{Marshall D.\ Newton}
\affiliation{Brookhaven National Laboratory, Chemistry Department, Box 5000, Upton, NY 11973-5000, United States}
 \email{newton@bnl.gov}
 \begin{abstract}
 Standard models of reaction kinetics in condensed materials rely on the Boltzmann-Gibbs distribution for the population of reactants at the top of the free energy barrier separating them from the products. While energy dissipation and quantum effects at the barrier top can potentially affect the transmission coefficient entering the rate preexponential factor, much stronger dynamical effects on the reaction barrier are caused by the breakdown of ergodicity for populating the reaction barrier (violation of the Boltzmann-Gibbs statistics). When the spectrum of medium modes coupled to the reaction coordinate includes fluctuations slower than the reaction rate, such nuclear motions dynamically freeze on the reaction time-scale and do not contribute to the activation barrier. Here we consider the consequences of this scenario for electrode reactions in slowly relaxing media. Changing electrode overpotential speeds  electrode electron transfer up, potentially cutting through the spectrum of nuclear modes coupled to the reaction coordinate. The reorganization energy of electrochemical electron transfer becomes a function of the electrode overpotential, switching between the thermodynamic value at low rates to the nonergodic limit at higher rates. The sharpness of this transition depends of the relaxation spectrum of the medium. The reorganization energy experiences a sudden drop with increasing overpotential for a medium with a Debye relaxation, but becomes a much shallower function of the overpotential for media with stretched exponential dynamics. The latter scenario characterizes electron transfer in ionic liquids.  The analysis of electrode reactions in room-temperature ionic liquids shows that the magnitude of the free energy of nuclear solvation is significantly below its thermodynamic limit. This result applies to reaction times faster than microseconds and is currently limited by the available dielectric relaxation data.         
 \end{abstract}

\maketitle

\section{Introduction}
The theory of nonadiabatic electron transfer at metal electrodes has been established in theoretical work by Marcus,\cite{Marcus:65} Hush,\cite{Hush:1968rr,Hush:99} and Levich\cite{Levich:1965wp} and extensively tested by a number of seminal experimental studies by Chidsey,\cite{Chidsey:91} Finklea,\cite{Finklea:2001ve} and Sav{\'e}ant.\cite{Saveant:2002em}  Nonadiabatic rate is calculated from the Golden-Rule perturbation theory for individual electronic transitions between the localized electronic state in solution and delocalized conduction states in the metal electrode.\cite{Chidsey:91,Hush:99,Gosavi:00,Gosavi:2001jh,Migliore:2012kx} The Golden-Rule expression is limited by a low magnitude of the electrode-reactant electronic coupling. As the coupling increases, two effects become important. First, at still relatively small coupling magnitudes, the dynamical solvent effect alters the pre-exponential factor of the rate constant for homogeneous electron transfer in solution\cite{Zusman:80,Sumi:86,Rips:1987qf,Yan:1988mz,Smith:1993dx} and for the electrode reaction.\cite{Morgan:1987bm,Chakravarti:1992ee,DMjec:94} At even higher electronic coupling, the barrier for electron transfer becomes affected in the regime known as adiabatic electrode reactions.\cite{Schmickler:86,DMjec:91,Schmickler:96,Hush:99,DMjcp2:09}     
  
While the theory of the solvent effect on electron transfer accounts for friction (dissipation) at the top of the barrier\cite{Kramers:1940wm,Smith:1993dx} and thus can be viewed as a nonequilibrium theory, it still assumes that the population of reactants near the barrier's top is given by the Boltzmann-Gibbs probability distribution. Therefore, all traditional theories mentioned above anticipate that the statistical configurations of the reactants in the reactants' well are assigned Boltzmann-Gibbs probabilities. The rate $k$ of an activated process is proportional to the equilibrium population of the activated state separated by the Gibbs energy barrier $\Delta G^\dag$ from the reactant state
\begin{equation}
k \propto e^{-\beta \Delta G^\dag},
\label{eq1}  
\end{equation}
where $\beta = 1/(k_\text{B}T)$ is the inverse temperature. 

\begin{figure}
\includegraphics[clip=true,trim= 0cm 0cm 0cm 0cm,width=6cm]{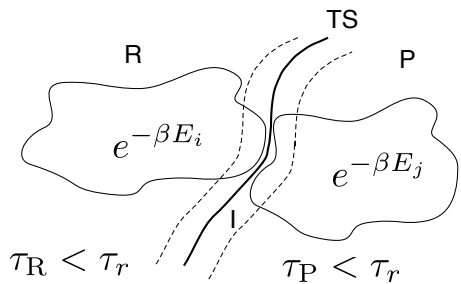}
\caption{Schematic representation of the barrier crossing in ergodic and nonergodic formulations of chemical kinetics. The transition state (TS) theory considers the reactive flux through a TS hypersurface separating the reactant (R) and product (P) parts of the phase space, each characterized by the corresponding Boltzmann-Gibbs distribution of statistical states. The stable states picture of Northrup and Hynes (NH)\cite{Northrup:1980hw} distinguishes an intermediate region (I) at the barrier top, where the medium dynamics are separately considered. The nonergodic kinetics can be applied to either TS or NH formalisms by lifting the approximation of fast relaxation within the reactant and product wells. Only the parts of the corresponding phase space (shown as closed regions) with sufficiently fast dynamics, $\tau_{s}<\tau_r=k^{-1}$, are included in the thermal bath  delivering particles to the top of the barrier. Boltzmann-Gibbs statistics is assumed within these constrained regions. }
\label{fig:0}  
\end{figure}

More detailed theories of the barrier crossing, following the original Kramers formalism,\cite{Kramers:1940wm} specifically consider the medium dynamics at the barrier top. Along these lines, the stable states picture proposed by Northrup and Hynes \cite{Northrup:1980hw} subdivides the system phase space into reactant and product stable regions, to which equilibrium Boltzmann-Gibbs distribution applies, and the intermediate region near the barrier top where nuclear dynamics are treated separately. This approach leads to memory effects in barrier crossing related to the flux-flux correlation functions for the particles entering and leaving the intermediate region (I in Fig.\ \ref{fig:0}). The separation into regions requires corresponding separation between the relaxation time $\tau_s$ inside each stable region  and the reaction time $\tau_r=k^{-1}$: $\tau_r\gg\tau_s$. Since relaxation within the reactant and product wells occurs fast on the reaction time-scale, one can use the Boltzmann-Gibbs statistics within the stable regions, and also apply it to calculate the flux-flux correlation functions according to standard recipes of the linear response approximation.\cite{Hansen:13} While the dynamics in the intermediate region are explicitly considered, the probability of reaching this region is still given by the Boltzmann-Gibbs distribution. This perspective is illustrated in Fig.\ \ref{fig:0}, where the picture of the dividing surface (TS) of the transition-state theory is confronted with the intermediate region of Northrup and Hynes. 

The nonergodic kinetics of chemical reactions\cite{DMacc:07,DMjcp1:09} makes the next step in introducing the medium dynamics into the rate of activated barrier crossing. The main difference of this perspective compared to the classical theories discussed above is that the limitation of fast relaxation within the reactant and product phase sub-space, $\tau_r\gg\tau_s$, is now lifted. The configurations allowed for sampling by the reactants and products are limited by the ability of the system to reach them on the time-scale of the reaction (closed areas in Fig.\ \ref{fig:0}). If some parts of the phase space require longer time to reach, those configurations do not contribute to the statistical averages used to calculate the reaction rate. The nonergodic kinetics thus adopts an ensemble view of insufficient sampling of the reactant and product phase space, instead of attempting to solve the entire dynamic problem involving nonequilibrium relaxation.\cite{Northrup:1980hw} The ensemble of stable states of Northrup and Hynes, from which the trajectories are pulled toward the barrier and which are assigned the Boltzmann-Gibbs weights, becomes depopulated, in addition to possible dissipative depopulation and recrossings at the top of the barrier considered in the Kramers and Northrup and Hynes models. 

\begin{figure}
\includegraphics[clip=true,trim= 0cm 1.5cm 0cm 0cm,width=8cm]{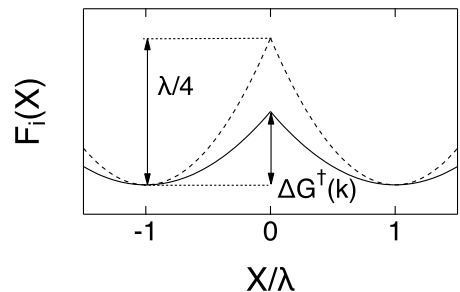}
\caption{Crossing of Marcus parabolas in the ergodic (dashed lines) and nonergodic (solid lines) scenarios for self-exchange electron transfer. The free energy surfaces $F_i(X)$ are drawn against the energy-gap reaction coordinate $X$. The crossing point $X=0$ defines the position of the activation barrier. The Marcus activation barrier is $\lambda/4$, where $\lambda$ is the thermodynamic reorganization energy. The nonergodic reorganization energy $\lambda(k)$ is lowered compared to $\lambda$, thus resulting in a lower activation barrier, $\Delta G^\dag(k)=\lambda(k)/4$. The barrier height becomes a function of the rate constant, which needs to be calculated by a self-consistent dynamic formalism,\cite{DMjpcb1:08} such as that exemplified by Eq.\ \eqref{eq2}. }  
\label{fig:00}
\end{figure}

This simple reasoning has significant implications for rates determined through Eq.\ \eqref{eq1}. Since the space of configurations is now restricted, the phase space contributing to the free energy profile along the reaction coordinate becomes restricted as well. The dynamics of the system now affect not only the barrier crossing event, but also the entire free energy profile of the reaction. The consequence of this perspective is best understood by considering the free energy profile of electron-transfer self exchange shown in Fig.\ \ref{fig:00}. Restricting the  available phase space lowers the reorganization energy of the reaction from its thermodynamic magnitude $\lambda$ to a nonergodic value $\lambda(k)$ (see below). This alteration propagates into the corresponding alteration of the entire free energy surface and leads to a lower activation barrier $\lambda(k)/4$. 

The reorganization energy $\lambda(k)$, and the barrier $\Delta G^\dag(k)=\lambda(k)/4$, are now functions of the rate constant. This is a general outcome of the nonergodic kinetics requiring that the free energy of activation $\Delta G^\dag(k)$ is affected by the medium dynamics through relative magnitudes of the reaction time $\tau_r=k^{-1}$ and $\tau_s$. Since the rate itself is dependent on the barrier height, it needs to be calculated by solving a self-consistent equation     
\begin{equation}
k \propto e^{-\beta \Delta G^\dag(k) } .
\label{eq2}  
\end{equation}  

Equation \ref{eq2} can be used directly to achieve a solution for $k$, as we do below, or within a more complex formalism. Theories of barrier crossing,\cite{Northrup:1980hw,Zusman:80,Sumi:86,Rips:1987qf,Yan:1988mz,Smith:1993dx} can be applied to calculate the self-consistent activation dynamics, including cases of non-exponential population kinetics when the rate constant is not uniquely defined.\cite{DMjpcb2:08} We do not address these issues here and instead, for the sake of simplicity of formulation, focus on the free energy profile for electron-transfer reactions. The rates are calculated from the well-established Golden-Rule approach.\cite{Levich:1965wp} We address electrode reactions in this picture as an important class of reactions for which the electrode potential becomes a control parameter which can potentially drive the reaction out of ergodicity.        

The conceptual framework of nonergodic kinetics clearly applies to electrode reactions in slowly relaxing media. Making the electrode  overpotential $\eta$ more negative in the case of reduction considered below leads to an exponential increase of the electrode reaction rate constant, $k=k_\text{el}(\eta)$, according to the Tafel law.\cite{Tafel:05,BardFaulkner01} With increasing reaction rate, the electrode reaction enters the window of nonergodic electrode kinetics when the time of electrode discharge $k_\text{el}(\eta)^{-1}$ and the characteristic relaxation time of the medium $\tau_s$ become approximately equal, 
\begin{equation}
\tau_s k_\text{el}(\eta)\simeq 1. 
\label{eq0}
\end{equation}
When this point is reached, the rate cannot be described by Eq.\ \eqref{eq1} any further and Eq.\ \eqref{eq2} should be used instead. Here, the overpotential $\eta$ is the deviation between the externally applied electrode potential $E$ and its equilibrium value $E_\text{eq}$: $\eta=E-E_\text{eq}$.\cite{BardFaulkner01}  

The electrode reactions present us with a unique capability to move, through a potential scan of the electrode, from ergodic kinetics, following the basic prescriptions of the transition-state theory,\cite{Eyring:80} to nonergodic kinetics with new rules for the reaction activation barrier affected by the medium dynamics. Here, we analyze the observable consequences of this general picture in application to electron transfer at the metal electrode.  We start with a generic case of a solvent relaxing by a single-exponential Debye law\cite{BoettcherII:73} and then consider a more complex case when a continuous manifold of Debye processes, described by stretched exponential dynamics,\cite{Ediger:96} is used for the solvent. For that purpose, we model  electrode reactions\cite{Fietkau:2006ia,Khoshtariya:2009mi,Fawcett:2011fp,Nikitina:2014ec} in room-temperature ionic liquids (RTILs).\cite{Arzhantsev:2007ds,Stoppa:2008be,Hunger:2009er,Weingrtner:2014ib}

\begin{figure}
\includegraphics[clip=true,trim= 0cm 1.5cm 0cm 0cm,width=8cm]{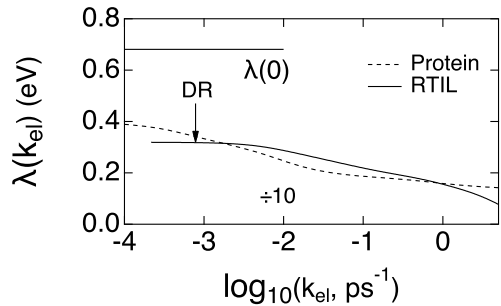}
\caption{Nonergodic reorganization energy $\lambda(k_\text{el})$ calculated for electrode electron transfer from the electrode to the ferrocene cation (solid lines) in $\mathrm{[bmim^+][PF_6^-]}$ (RTIL). The horizontal line marks the thermodynamic reorganization energy, $\lambda(0)=0.68$ eV, calculated with $\epsilon(\omega)=\epsilon_s\to\infty$ applied to the longitudinal susceptibility in Eq.\ \eqref{eq26}. The dashed line shows results of molecular dynamics simulations for the reorganization energy of bacterial $\mathrm{bc_1}$ complex calculated from 13 $\mu$s  simulations.\cite{DMjcp1:15} The reorganization energy for protein electron transfer is scaled down by a factor of 10 to fit the scale of the plot. The vertical arrow indicates the lowest frequency accessible by the dielectric relaxation (DR) spectroscopy.\cite{Stoppa:2008be} 
}
\label{fig:1}
\end{figure}

RTILs is a specific case of a broad list of materials characterized by stretched exponential dynamics and spanning several orders of magnitude in their relaxation times. Proteins is another example,\cite{Hong:2011qf,Khodadadi:2015jp,Mondal:2017gf} along with a large number of glass formers.\cite{Ediger:96}  The dependence of the reorganization energy on the rate constant $\lambda(k_\text{el})$ calculated here for RTILs is very shallow (Fig.\ \ref{fig:1}). Importantly, the reorganization energy does not reach its thermodynamic value in the entire range of $k_\text{el}$ considered here. The accessible range of rate constants is limited by the current experimental window of dielectric spectroscopy.\cite{Stoppa:2008be} The dependence of $\lambda(k_\text{el})$ calculated here turns out to be similar to that found previously for proteins\cite{DMjcp1:15} (Fig.\ \ref{fig:1}). This similarity suggests a common phenomenology for these types of media, which is quite distinct from a much sharper nonergodic crossover of the reorganization energy found for reactions in media with Debye relaxation.\cite{DMjpcb1:06}

\section{Rate of electrode electron transfer }
Essentially all complex media show complex dynamics, and hence deviations of their time-correlation functions from the simple one-exponential (Debye) form. This phenomenology is usually understood in two ways, which are not necessarily opposing each other. One can view the non-exponential kinetics as either a reflection of a large number of Debye decay relaxation processes with their characteristic time-scales\cite{Richert:02,Richert:2014wa}  or a continuous distribution of Debye decays resulting, mathematically, in complex dynamic susceptibilities and non-Debye functionalities for the dynamic response functions.\cite{Bagchi:91} The former mechanism is often linked to dynamic heterogeneity\cite{Richert:02} and the latter mechanism can be related to intrinsically complex dynamics of condensed materials,\cite{Hansen:13} not necessarily involving spatial separation of dynamically distinct regions. The actual mechanism of appearance of many relaxation times is not central for the description of reactions in such complex media. The main feature from the viewpoint of the reaction dynamics is that nonequilibrium fluctuations of the nuclear modes coupled to the reaction need to be represented by several processes, some fast and some slow. 

This perspective offers a possibility of dynamical freezing: that is, some of the modes cannot relax on the time-scale of the reaction.\cite{Chen:96,Gorlach:95,Richert:2000wq,Goes:02,DMacc:07} The typical relaxation phenomenology of polar liquids and of many complex media (such as proteins\cite{Lakowicz:2000jn,DMjcp2:13,Qin:2017kk}) involves two basic components: fast ballistic relaxation\cite{Jimenez:94,BagchiGayathri:99} and much slower collective relaxation, which can be multiexponential or characterized by complex dynamics (such as stretched exponential\cite{Ediger:96,Ediger:00,Richert:02}). For charge-transfer transitions, the relevant dynamic property is the Stokes-shift time autocorrelation function\cite{Zwan:85,Carter:91,Smith:1993dx,Reynolds:96}
\begin{equation}
C_X(t) = \langle \delta X(t) \delta X(0)\rangle . 
\label{eq2-1}  
\end{equation}

The correlation function $C_X(t)$ projects the nuclear dynamics of the medium on the collective variable corresponding to the energy gap $X(t)$ (as shown in Fig.\ \ref{fig:00}) between electronic energies before and after the electronic transition, $\delta X(t)=X(t) -\langle X\rangle $. For electrochemical reactions, the initial state of the cathodic process is the electron localized at the energy state $\epsilon$ in the conduction band of the metal and the reactant in the oxidized state with the energy $E^\text{Ox}$. The initial energy is $\epsilon+E^\text{Ox}$. The final state, after the electron transfer, has the energy of the reduced reactant $E^\text{Red}$. The energy gap is therefore $X=E^\text{Red}-E^\text{Ox}-\epsilon$. Since the energy level $\epsilon$ in the metal is screened from the medium fluctuations, one can put $\delta X(t)=  \delta E^\text{Red}(t) -\delta E^\text{Ox}(t)$ in Eq.\ \eqref{eq2-1}. The time dependence of the energy gap fluctuations is caused by thermal agitation of the nuclear modes of the solvent medium, and the stationary value $C_X(0)$ defines the equilibrium (Gibbs ensemble) reorganization energy $\lambda$\cite{Mukamel:95} 
\begin{equation}
\lambda = \beta C_X(0)/2 .  
\label{eq2-3}
\end{equation}

\begin{figure}
\includegraphics[clip=true,trim= 0cm 1.5cm 0cm 0cm,width=8cm]{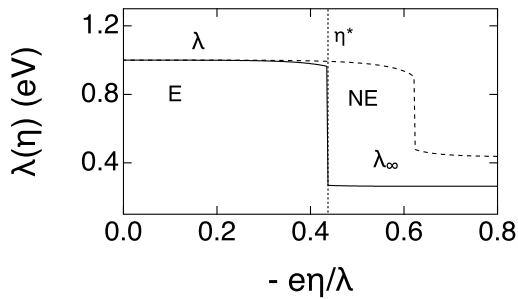}
\caption{$\lambda(\eta)$ from Eq.\ \eqref{eq3} calculated at $\kappa_s=2\tau_s \Delta /\hbar$ (Eq.\ \eqref{eq7}) equal to $2.0$ (dashed line) and 8.0 (solid line). $\lambda=\lambda_f+\lambda_s$ is the thermodynamic (ergodic) reorganization (free) energy and $\lambda_\infty$ is the nonergodic reorganization energy in the fast reaction limit given by Eq.\ \eqref{eq4}. The dashed vertical line marks the crossover between the region of ergodic (E) and nonergodic (NE) reactions. The crossover overpotential $\eta^*$ is calculated from Eq.\ \eqref{eq8}. The fast, $\lambda_f=0.2$ eV, and slow, $\lambda_s=0.8$ eV, components of the reorganization energy are used in the calculations. The electrode overpotential is scaled with the thermodynamic reorganization energy $\lambda=\lambda_f+\lambda_s$.     
}
\label{fig:2}  
\end{figure}

In contrast to this thermodynamic reorganization energy, nonergodic reorganization energy arises in theories of activated transitions constraining the space of configurations available to the system. The constraint is dynamical, demanding that no relaxation process slower than the reaction time contributes to the statistical averages specifying the activation barrier\cite{DMjcp2:05,DMjcp1:09,DMjpcm:15} (Fig.\ ref{fig:0}). Such constraints are mathematically represented by the frequency filter $\omega>k_\text{el}$ when the nonergodic reorganization energy is expressed in terms of the time Fourier transform $\bar C_X(\omega)$ of the Stokes correlation function. One obtains
\begin{equation}
  \lambda(k_\text{el}) = \int_{k_\text{el}}^\infty \chi\strut_X''(\omega)\ d\omega/(\pi\omega),
\label{eq2-2}  
\end{equation}
where $\chi\strut_X''(\omega)=\omega \beta \bar C_X(\omega)/2$ is the imaginary part of the complex-valued Stokes-shift susceptibility following from the standard rules of the linear-response approximation\cite{Hansen:13} formulated in terms of the collective variable $X(t)$. Its imaginary part, a loss function, determines the rate of loss of energy supplied to match a given value of $X$ (by say an optical excitation) into the thermal bath.

If the long-time decay of the Stokes-shift correlation function is single exponential, as we assume here at the first step of our modeling, one obtains\cite{DMjcp1:09} from Eq.\ \eqref{eq2-2} 
\begin{equation}
\lambda(\eta) = \lambda_f +  (2\lambda_s/\pi) \mathrm{arccot}\left[k_\text{el}(\eta)\tau_s\right]. 
\label{eq3}   
\end{equation}    
Here, $\lambda_f$ is the reorganization energy of the fast modes, $\lambda_s$ is the reorganization energy  of the slow modes, and $\tau_s$ is the characteristic relaxation time of the slow component of $C_X(t)$. 

Equation \eqref{eq3} is written in the form applicable to electrochemistry to stress the dependence of the reorganization energy on the electrode overpotential $\eta$ through the  corresponding dependence of the the rate of electrode electron transfer $k_\text{el}(\eta)$. Note that $\lambda=\lambda_f+\lambda_s$ is a thermodynamic free energy in the standard models\cite{MarcusSutin} as given by Eq.\ \eqref{eq2-3}. It does not depend of the electrode overpotential. However, when the reaction time becomes comparable in magnitude to the slow relaxation time $\tau_s$, the reorganization energy gains a dependence on the overpotential. A representative calculation is shown in Fig.\ \ref{fig:2}. When the reaction rate exceeds the rate of slow medium relaxation, the reorganization energy drops as a function of $\eta$ from its equilibrium value $\lambda=\lambda_f+\lambda_s$ to the reorganization energy of the fast reaction  
\begin{equation}
\lambda_\infty=\lambda_f + (2\lambda_s/\pi)\mathrm{arccot}[\kappa_s],
\label{eq4}  
\end{equation}
where the parameter 
\begin{equation}
\kappa_s = 2\tau_s\Delta/\hbar  
\label{eq5} 
\end{equation}
gives the ratio of the rate of activationless electrode reaction (the rate of electron tunneling) to the rate of medium relaxation. Here, $\Delta =\pi V^2\rho_F$,\cite{Schmickler:86,DMjec:91,Schmickler:96,Gosavi:00,DMjcp2:09} $V$ is the electron-transfer matrix element between the electronic state on the reactant and a single electronic state in the metal and  $\rho_F$ is the density of states at the metal's Fermi level.

The reorganization energy $\lambda_\infty$ loses the status of the thermodynamic free energy and becomes a nonergodic energy parameter quantifying reorganization of the nuclear subsystem on the time-scales shorter than the electron tunneling rate $2\Delta/\hbar$ (maximum (activationless) nonadibatic rate\cite{Hush:99,Hale:1968ju}). In the general case, the reorganization energy $\lambda(\eta)$ enters the rate of  nonadiabatic electrode reaction, usually obtained by integrating individual nonadiabatic transitions over the Fermi-distributed conduction electrons in the metal\cite{Levich:1965wp,Chidsey:91,Hush:99,Migliore:2012kx,Oldham:2011fk,Newton:2007fk} 
\begin{equation}
k_\text{el}(\eta)= \frac{\Delta}{\hbar} \left(\frac{\beta}{\pi\lambda(\eta)}\right)^{1/2}\int_{-\infty}^\infty e^{-\beta \Delta G^\dag(\eta-\epsilon/e)} f_F(\epsilon) d\epsilon ,
\label{eq5-1}
\end{equation}
where $f_F(\epsilon)=[\exp(\beta\epsilon)+1]^{-1}$ is the Fermi population function
and the activation barrier $\Delta G^\dag(\eta)$ is the standard result of the Marcus-Hush (MH) theory
\begin{equation}
\Delta G^\dag(\eta) = \frac{(\lambda(\eta) + e\eta)^2}{4\lambda(\eta)} .
\label{eq5-2}
\end{equation}
 
If the Fermi population $f_F$ is taken at zero temperature the resulting relation is\cite{Hale:1968ju}
\begin{equation}
k_\text{el}(\eta)= \frac{\Delta}{\hbar}\ \mathrm{erfc}\left([\beta\Delta G^\dag(\eta)]^{1/2}\right) ,   
\label{eq6}
\end{equation}
where $\mathrm{erfc}(x)$ is the complementary error function.\cite{Abramowitz:72} From Eq.\ \eqref{eq6}, $k_\text{el}(\eta)\to 2\Delta /\hbar$ at $-e\eta\gg \lambda(\eta)$. This limit corresponds to an activationless electrode electron transfer, the rate of which is used in defining the parameter $\kappa_s$, as discussed above in connection with Eq.\ \eqref{eq5}.   

While Eq.\ \eqref{eq6} gives a reasonable approximation for the rate, a more accurate preexponential factor is calculated by integrating over the Fermi distribution at a finite temperature. The procedure suggested by Marcus and co-workers\cite{Gosavi:00,Gosavi:2001jh} involves neglecting or Taylor expanding the Gaussian function $\exp[-\beta\epsilon^2/(4\lambda)]$ in the $\epsilon$-integral in Eq.\ \eqref{eq5-1}. Merely dropping this term produces an analytical form 
\begin{equation}
k_\text{el}(\eta)= \frac{\Delta}{\hbar} \left(\frac{\pi}{\beta\lambda(\eta)}\right)^{1/2}\sec \left(\frac{0.85\pi e\eta}{2\lambda(\eta)}\right) e^{-\beta \Delta G^\dag(\eta)},  
\label{eq6-1}
\end{equation}
which is valid under the condition of $e|\eta|<\lambda(\eta)$. We have empirically introduced the factor 0.85 in the argument of $\sec(x)=[\cos(x)]^{-1}$  to account for the terms dropped from the expansion of $\exp[-\beta\epsilon^2/(4\lambda)]$ in Eq.\ \eqref{eq5-1}. It brings Eq.\ \eqref{eq6-1} into good agreement with numerical integration. We note that the rate calculated here refers to a reactant placed at a specific distance from the electrode and has the units of inverse time. It can therefore be directly used for reactants adsorbed at the electrode. Alternatively, when mass transport becomes involved, the rate at a given distance is an input for the corresponding diffusional kinetics model\cite{ComptonBanks} producing the electrochemical rate in units of length/time.\cite{DMjec:94} Our examples below apply to adsorbed redox species to avoid complications from mass transport.    

\begin{figure}
\includegraphics[clip=true,trim= 0cm 1.5cm 0cm 0cm,width=8cm]{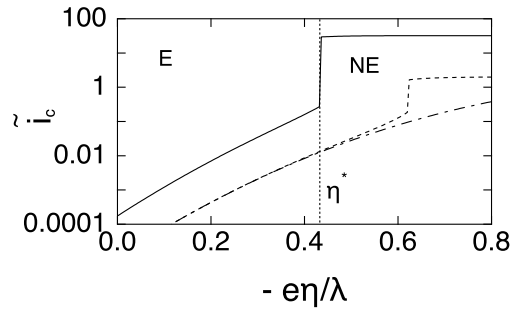}
\caption{Scaled cathodic current $\tilde i_c$ (logarithmic scale, Eq.\ \eqref{eq7}) vs the electrode overpotential $\eta$. The calculations are done for $\kappa_s=2\tau_s \Delta /\hbar$ (Eq.\ \eqref{eq5}) equal to $2.0$ (dashed line) and 8.0 (solid line), $\lambda=\lambda_f+\lambda_s=1.0$ eV. The dash-dotted line ($\kappa_s=2.0$) shows the standard result for nonadiabatic electrode electron transfer, which is obtained upon substitution $\lambda(\eta)\to\lambda$ in Eq.\ \eqref{eq6}. The vertical dotted line indicates the cross-over overpotential $\eta^*$ calculated from Eq.\ \eqref{eq8} and separating the ergodic region (E) from the nonergodic region (NE).  
}
\label{fig:3}  
\end{figure}

\section{Stationary electrode current} 
We want to establish how nonergodicity of the activation barrier for electrode electron transfer affects the observable electrode current. We will consider here the simplification of the electrochemical setup in which the reactants are assumed to be immobilized at the surface of the electrode\cite{Chidsey:91,Yue:2006wo,Newton:2007fk} with the total surface concentration $\Gamma_t=\Gamma_\text{Ox}+\Gamma_\text{Red}$ based on the surface concentrations of the oxidized, $\Gamma_\text{Ox}$, and reduced, $\Gamma_\text{Red}$, forms of the reactant. This setup thus eliminates the need to consider the diffusive mass transport.\cite{ComptonBanks}   

In order to gain physical insights into how non-ergodicity modifies the electrode kinetics, we start with a simplified and somewhat unrealistic model in which we consider the cathodic process with a fixed concentration of the oxidized form $\Gamma_\text{Ox}=\Gamma_t$. We thus consider the reduction process $A^+ + e^-\to A$. Since the model is symmetric in respect to the overpotential, the oxidative branch follows from the sign flip, $\eta\to -\eta$. We calculate the dimensionless current 
\begin{equation}
\tilde i_c = i_c\tau_s/(A\Gamma_t) = k_\text{el}\tau_s  
\label{eq7}
\end{equation}
from the total cathodic current $i_c$ passing through the electrode area $A$. 

Figure \ref{fig:3} shows $\tilde i_c$ from the rate constant determined by combining Eqs.\ \eqref{eq3} and \eqref{eq6}. The main result from these calculations is the identification of the overpotential of the dynamical crossover $\eta^*$ determined by the condition (empirically found from Eqs.\ \eqref{eq4} and \eqref{eq6})
\begin{equation}
\tau_s k_\text{el}(\eta^*) \simeq 0.7  
\label{eq8}  
\end{equation}
at which an approximately exponential (Tafel law\cite{Tafel:05,ComptonBanks}) dependence of the current on overptential switches to the saturation limit of activationless electron tunneling between the electrode and the reactant. The MH model for nonadiabatic electron transfer (dash-dotted line in Fig.\ \ref{fig:3}) reaches the activationless regime at $-e\eta=\lambda$ (as follows from the rate constant in Eq.\ \eqref{eq6} upon the substitution $\lambda(\eta)\to \lambda$). It requires a higher cathodic overpotential to reach the activationless reaction than in the nonergodic model.      

\begin{figure}
\includegraphics[clip=true,trim= 0cm 1.5cm 0cm 0cm,width=8cm]{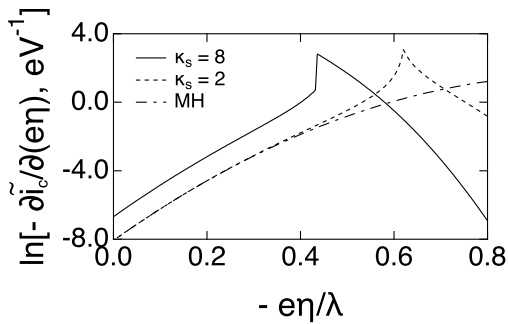}
\caption{$-d\tilde i_c/d(e\eta)$ calculated for nonergodic electrode kinetics: $\kappa_s$ (Eq.\ \eqref{eq4}) is equal to 8.0 (solid line) and 2.0 (dashed line). The dash-dotted line shows the result of the Marcus-Hush theory (MH, Eqs.\ \eqref{eq9} and \eqref{eq10}).  
}
\label{fig:4}
\end{figure}

More detailed information about the distribution of the reactant energy levels in the medium can be achieved by taking the derivative of the current over the overpotential.\cite{Becka:1992mz,Terrettaz:1996gv,Miller:95} Within the MH model this observable directly probes the Gaussian distribution of the oxidized state. One obtains from Eq.\  \eqref{eq6}
\begin{equation}
-d \tilde i_c/d(e\eta) = \kappa_s P_G(\eta), 
\label{eq9}  
\end{equation}
where
\begin{equation}
P_G(\eta) = \left[4\pi\lambda k_\text{B}T \right]^{-1/2} \exp\left[ -\beta \Delta G^\dag(\eta)\right] 
\label{eq10}  
\end{equation}
is the Gaussian function (see Eq.\ \eqref{eq5-2} at $\lambda(\eta)=\lambda$) of the driving force $-e\eta$ in the MH theory. In the nonergodic theory, one has to include the derivative of $\lambda(\eta)$ with respect to $\eta$, which results in a slightly more complex relation.

The non-Gaussian shape of the nonergodic current derivative arises from the combination of two essentially Gaussian functions in one plot. The reorganization energy is high at low overpotentials and one sees the rising wing of the Gaussian probability function enroute to achieving the maximum at $-e\eta = \lambda = \lambda_g+\lambda_s$ (cf.\ dashed and dash-dotted lines in Fig.\ \ref{fig:4}). However, the reorganization energy drops due to nonergodic constraints before this limit is reached (Fig.\ \ref{fig:2}). Correspondingly, further increasing $|\eta|$ (making it more negative) displays the decaying wing of the Gaussian function characterized by $\lambda_\infty$ (Eq.\ \eqref{eq4}). What is seen in Fig.\ \ref{fig:4} is an overlap, in one plot, of these two Gaussian curves.

\begin{figure}
\includegraphics[clip=true,trim= 0cm 1.5cm 0cm 0cm,width=8cm]{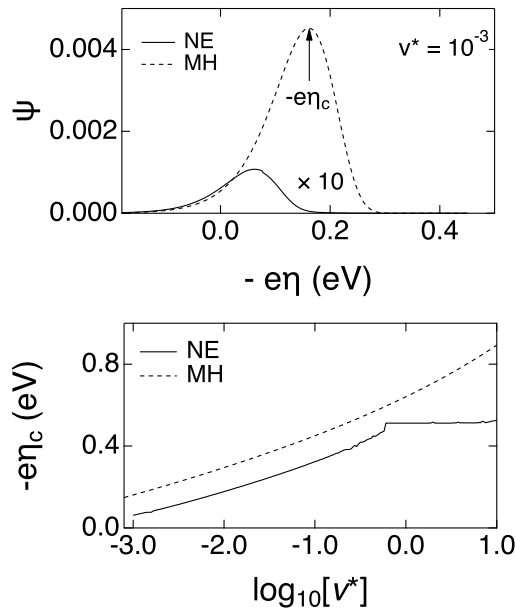}
\caption{Cathodic cyclic voltammetry wave (upper panel, Eq.\ \eqref{eq11-1}) and the dependence of the overpotential of the cathodic wave maximum, $\eta_c$,  on the scaled scanning rate $v^*$ (Eq.\ \eqref{eq13}, lower panel). The results of calculations with the Marcus-Hush (MH) and nonergodic (NE) theories are compared. The current in the NE model is multiplied by a factor of 10 in the upper panel to bring it to the scale of $\psi$ (defined by Eq.\ \eqref{eq11-1}) in the MH model. The parameters used in the calculations are: $\kappa_s=4$, $\lambda_f=0.2$ eV, and $\lambda_s=0.8$ eV.   }
\label{fig:5}
\end{figure}

\section{Cyclic voltammetry } 
The calculations performed in the previous section represent the idealized situation when the surface concentration of the oxidized state is not affected by the passing of the cathodic current. Here, we turn to a more realistic description in terms of linear sweep cyclic voltammetry (CV), where the surface concentration changes when the overpotential is altered with the scan rate $v$: $\eta=\eta_m-vt$.  Thus the cathodic sweep runs from $\eta_m$ to $-\eta_m$ with the scan rate $v$.   

The total electrode current $i=i_c-i_a$ is composed of the cathodic, $i_c$, and anodic, $i_a$, currents passing through the area $A$ under the applied overpotential. One can define, following Laviron,\cite{Laviron79} the scaled current 
\begin{equation}
  \psi=\frac{i}{\beta e^2 v A \Gamma_t}. 
  \label{eq11-1}
\end{equation}

 The equation for the scaled current is given in terms of the surface mole fractions of the oxidized, $x_\text{O}=\Gamma_\text{O}/\Gamma_t$, and reduced, $x_\text{R}=1-x_\text{O}$, adsorbates; $\Gamma_t$ is the total surface concentration. The equation for the current is\cite{Honeychurch:1999uq}
\begin{equation}
\psi = (k_\text{O}/v^*) x_\text{O} - (k_\text{R}/v^*)(1-x_\text{O}) .
\label{eq12}  
\end{equation}
Here,
\begin{equation}
v^* = \beta e v\tau_s  
\label{eq13}  
\end{equation}
is the dimensionless scan rate. Further, the scaled rates for the oxidation and reduction reactions in Eq.\ \eqref{eq12} can be calculated either from integration over the Fermi distribution of electron states in Eq.\ \eqref{eq5-1} or from the step-wise approximation of the Fermi distribution, leading to Eq.\ \eqref{eq6}. Perturbative models are also possible,\cite{Gosavi:00,Gosavi:2001jh,Migliore:2012kx} but they produce only minor corrections to the rate preexponential factor. In the case of $T=0$ assumed for the distribution of electrons in the metal, one gets 
\begin{equation}
k_\text{O,R}(\phi) = \frac{\kappa_s}{2}\mathrm{erfc}\left(\frac{\sqrt{\beta\lambda(\phi)}}{2}\pm \frac{\phi}{\sqrt{\beta\lambda(\phi)}}\right) \label{eq14}  
\end{equation}
where $k_\text{O,R}=\tau_s k_\text{el}$ are scaled rate constants, with ``$+$'' and ``$-$'' applied to ``O'' and ``R'', respectively, and $\phi = \beta e\eta$. Equation \eqref{eq6-1} is not very convenient for the modeling of cyclic voltammetry because of the limited range of overpotentials to which it applies.

The solution for $x_\text{O}(\phi)$ is given as\cite{Honeychurch:1999uq}
\begin{equation}
\begin{split}
x_\text{O}(\phi) &= e^{\frac{1}{v^*}\int_{\phi_m}^{\phi} k_t dz }\\
                 &- \frac{1}{v^*}\int_{\phi_m}^{\phi} dz k_\text{R}(z) e^{\frac{1}{v^*}\int_z^{\phi} k_t dy  } ,
\end{split}
\label{eq15}  
\end{equation}
where $k_t = k_\text{O}+k_\text{R}$. We have additionally assumed that the sweep amplitude is large enough such that only the oxidized form is present at the electrode at $\eta=\eta_m$: $x_\text{O}(\eta_m)=1$. With this initial condition, Eq.\ \eqref{eq15} is the solution of the following kinetic equation 
\begin{equation}
\tau_s \dot x_\text{O} = - k_\text{O} x_\text{O} +  k_\text{R} \left(1 - x_\text{O}\right) .
\label{eq16}  
\end{equation}
  
The results of these calculations are shown in Fig.\ \ref{fig:5}. The upper panel shows the cathodic CV peak at the scaled scanning rate $v^*=10^{-3}$ in the MH and nonergodic kinetics. The nonergodic reorganization energy produces a lower barrier and a faster rate. The population of the oxidized state of the reactant is depleted faster, with the resulting shift of the peak potential to a lower value and a corresponding decrease in its amplitude. Large scanning rates eventually result in leveling off of the peak position as a function of $v^*$ (solid line in the lower panel in Fig.\ \ref{fig:5}). However, reaching this regime requires very low relaxation times $\tau_s$ in Eq.\ \eqref{eq13}. For most practical systems, only $v^*\ll 1$ can be realized.

\section{Electrode reactions in ionic liquids}  
The standard Marcus model based on dipolar polarization of the bath predicts that the medium (solvent) reorganization energy is proportional to the Pekar factor, $c_0=\epsilon_\infty^{-1}-\epsilon_s^{-1}$. While the optical dielectric constant $\epsilon_\infty$ is well defined for all media, the static dielectric constant $\epsilon_s$ is expected to diverge at $\omega\to 0$ for RTILs due to their intrinsic conductivity.  This notion implies that a sample of conducting material placed in a constant external electric field must develop surface charges, and a corresponding macroscopic dipole moment, in order to render the field inside the conductor equal to zero.\cite{Landau8} This perfect screening of the external field is what is meant by an infinite static dielectric constant,\cite{Lynden-Bell:2007pd,LyndenBell:2007ff,Hansen:13} which cannot be ``corrected'' by the conductivity loss\cite{Weingrtner:2014ib} in the imaginary part of the frequency-dependent dielectric function reported by dielectric spectroscopy. The ``static'' dielectric constant of $\epsilon_s\simeq 10-25$, often cited in the literature ($\sim 15-35$ for protic RTILs\cite{Sonnleitner:2015fla,Weingrtner:2014ib}), mostly refers to the GHz domain of frequencies\cite{Stoppa:2008be,Hunger:2009er} (down to 0.2 GHz in Ref.\ \onlinecite{Sonnleitner:2015fla}). Measurements in the kHz domain are dominated by electrode polarization effects\cite{Leys:2008il,Serghei:2009dl} and do not produce reliable data at high temperatures; dielectric data down to $10^{-2}$ Hz are available near the glass transition.\cite{Ito:2007tw} 
  
These comments are meant to emphasize that polar response in ionic liquids has to refer to a specific frequency window. Dielectric spectroscopy covers the time-scales relevant for solvation dynamics and for many redox reactions. However, since the long-time dynamics are dispersive and cover 2--3 orders of magnitude in time-scales,\cite{Arzhantsev:2007ds,Stoppa:2008be} dynamic freezing of a part of the spectrum can always be an issue. Therefore, nonergodic effects are potentially important and RTILs are a primary target for the nonergodic theory of electron transfer. To understand consequences of nonergodic effects for electrode reactions, we perform here calculations modeling electrode reduction of ferrocene cation\cite{Fietkau:2006ia} in $\mathrm{[bmim^+][PF_6^-]}$.   

Ionic liquids are obviously different from dipolar molecular solvents,\cite{Wang:2007iy,Fayer:2014et} but still match in many respects dielectric properties of molecular liquids.\cite{Lynden-Bell:2007pd,LyndenBell:2007ff,Shim:2009jf,Mladenova:2016gu} Polar response of an ionic liquid is mostly due to translational motions of the ions. Rotations of molecular dipoles $m$ (mostly belonging to cations\cite{Kashyap:2008fe}) were considered as an alternative mechanism of polar response, but are now viewed as a minor contribution to solvation as a result of recent computer simulations.\cite{Roy:2012gp,Terranova:2013dj} Our goal here is to establish a mapping of charge density fluctuations onto fluctuations of a dipolar polarization field, thus connecting calculations of rates to dielectric measurements. 

\begin{figure}
\includegraphics[clip=true,trim= 0cm 0cm 0cm 0cm,width=3cm]{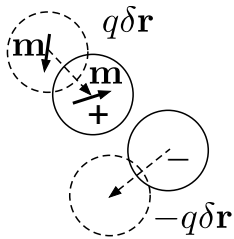}
\caption{Schematic representation of electrostatic fluctuations in RTILs. Translations of ions from initial positions (dashed circles) to the final positions (solid circles) are characterized by dipole moments $\pm q\delta\mathbf{r}_j$, rotations of molecular dipoles produce fluctuations of the molecular dipoles $\mathbf{m}_j$. The dipole moment is defined in respect to the center of mass of the corresponding molecule. These two types of thermal motions combined lead to a dipolar  polarization field characterizing electrostatic fluctuations in the RTIL.    
}
\label{fig:6}  
\end{figure}

A structural fluctuation of an ionic liquid results in a fluctuation of charge density, which can be mapped on a dipolar polarization field. This can be illustrated by assigning dipoles to ionic translations (Fig.\ \ref{fig:6}). The resulting dipolar field represents a fluctuating charge density by its divergence, $\delta\rho^Z=-\nabla\cdot \mathbf{P}$. The charge  density in reciprocal space relates to the longitudinal projection of the polarization field, $\delta\rho_\mathbf{k}^Z=ik P_{L}(\mathbf{k})$. Therefore, charge density fluctuations produced by ionic translations can be mapped on the longitudinal polarization field, and the longitudinal dipolar response can be used to model solvation.

If the solute interacts with the charges of the ionic liquid via the electrostatic potential $\phi_0(\mathbf{r})$, the Stokes shift susceptibility can be calculated from the linear response approximation.\cite{Hansen:13} The result is conventionally written in reciprocal space
\begin{equation}
\chi\strut_X(\omega) = \tfrac{1}{2}\int \frac{d\mathbf{k}}{(2\pi)^3} |\phi_{0\mathbf{k}}|^2 k^2\tilde\chi\strut_L(k,\omega) .
\label{eq17}
\end{equation}
Here, a factor $1/2$, which does not appear for the response of the bulk,\cite{DMmp:93,DMjcp1:17} accounts for the fact that real-space integration is performed over half of the entire space since the space occupied by the polar liquid is restricted by the electrode. The tilde for $\tilde \chi\strut_L$ in Eq.\ \eqref{eq17} and for the related functions below specifies the Laplace-Fourier transform\cite{Hansen:13} (Laplace transform with the imaginary variable) of the corresponding time-dependent function. Both $\chi\strut_X(\omega)$ and $\tilde \chi\strut_L(k,\omega)$ are therefore complex-valued functions. 

The longitudinal susceptibility can be expressed in terms of the longitudinal nonlocal dielectric function of the medium $\epsilon\strut_L(k,\omega)$
\begin{equation}
\tilde\chi\strut_L(k,\omega)=\frac{1}{4\pi}\left(1-\frac{1}{\epsilon\strut_L(k,\omega)}\right)\left(\frac{\epsilon_\infty+2}{3\epsilon_\infty}\right)^2.
\label{eq18}
\end{equation}
The factor containing the high-frequency dielectric constant $\epsilon_\infty$ in this relation accounts for the screening of the permanent charges by the induced dipoles described by the Lorentz local field.\cite{DMjcp1:17}

The longitudinal susceptibility function follows from the standard relations of the linear response theory\cite{Hansen:13}  
\begin{equation}
\tilde \chi\strut_L(k,\omega) =\chi\strut_L(k)\left[1+i\omega\tilde\Phi_L(k,\omega)\right].   
\label{eq21} 
\end{equation}
Here, $\tilde\Phi^L(k,\omega)$ is the Laplace-Fourier transform of the normalized correlation function of the longitudinal (nuclear) polarization field representing electrostatic fluctuations of the solvent  
\begin{equation}
\Phi_L(k,t) = \frac{\langle P_L(\mathbf{k},t)P_L(-\mathbf{k},0)\rangle }{\langle |P_L(\mathbf{k},0)|^2\rangle} .  
\label{eq22}
\end{equation}

The function $\tilde \Phi_L(k,\omega)$ can be alternatively written in terms of the memory function\cite{Boon:80,Hansen:13} $M_L(k,\omega)$ as 
\begin{equation}
  \tilde\Phi_L(k,\omega) = \left[-i\omega + M_L(k,\omega)\right]^{-1} .
  \label{eq23}
\end{equation}
Equation \eqref{eq23} can be viewed as a definition of the memory function, which generally is not directly related to observable properties. However, for the problem of charge fluctuations, the memory function is related to the observable property of longitudinal conductivity $\sigma_L(k,\omega)$
\begin{equation}
M_L(k,\omega) = 4\pi \sigma\strut_L(k,\omega). 
\label{eq19}
\end{equation}
Longitudinal conductivity connects the longitudinal electric current $\mathbf{k}\cdot \mathbf{J}_\mathbf{k}$ to the longitudinal Maxwell field $\mathbf{k}\cdot\mathbf{E}_\mathbf{k}$ as follows: $\mathbf{k}\cdot \mathbf{J}_\mathbf{k} = \sigma_L(k,\omega) \mathbf{k}\cdot\mathbf{E}_\mathbf{k}$, where both the current and the field oscillate with the frequency $\omega$.

A general functionality for $M_L(k,\omega)$ is unknown and requires extensive calculations even for model systems.\cite{Wilke:1999iq} The simplest approximation is to neglect the frequency dependence altogether. This representation conveniently connects the general formalism  with the limiting case of Debye relaxation.\cite{BoettcherII:73}  One obtains from Eqs.\ \eqref{eq2-2} and \eqref{eq23} 
\begin{equation}
\lambda(k_\text{el}) = \frac{1}{\pi}  \int \frac{d\mathbf{k}}{(2\pi)^3} \left|\phi_{0\mathbf{k}}\right|^2 k^2 \chi\strut_L(k)  \mathrm{arccot}[k_\text{el}\tau\strut_L(k) ] . 
\label{eq24}
\end{equation}
Instead of a single Debye relaxation time $\tau_s$ in Eq.\ \eqref{eq3}, a continuous distribution of relaxation times $\tau_L(k)=[M_L(k)]^{-1}$  enters the polar response with the weights specified by the static longitudinal susceptibility $k^2\chi\strut_L(k)$. The contribution of the slow relaxation modes to the integral is reduced by the function $\mathrm{arccot}[k_\text{el}/M_L(k) ]$, which tends to zero when the reaction rate is faster than the corresponding relaxation time. 

The next level of approximation for the memory function allows one to circumvent some modeling difficulties by connecting to dielectric experiment incorporating the complex dynamics. It is achieved by factorizing  $M_L(k,\omega)$ into functions of $k$ and $\omega$\cite{Munakata:75,Fried:90} 
\begin{equation}
  M_L(k,\omega) = m_L(\omega) \frac{\langle|\dot P_L(\mathbf{k},0)|^2 \rangle }{\langle|P_L(\mathbf{k},0)|^2 \rangle } .
  \label{eq25}
\end{equation}
This factorization allows one to determine $m_L(\omega)$ by satisfying the connection to the $k=0$ limit in Eq.\ \eqref{eq18}. The final result is
\begin{equation}
\tilde\chi\strut_L(k,\omega)= \chi_L(k)\left[1+ \frac{\chi\strut_L(k)}{\chi\strut_L(0)} \frac{\epsilon_s-\epsilon(\omega)}{\epsilon_s(\epsilon(\omega)-1)} \right]^{-1} ,
\label{eq26}
\end{equation}
where $\epsilon_s$ is the static dielectric constant. We put $\epsilon_s\to \infty$ in the calculations presented below. This limit, taken in Eq.\ \eqref{eq26}, implies that the frequency dependence of the longitudinal response is determined by the dielectric modulus\cite{Ito:2007tw} $M(\omega)=[\epsilon(\omega)]^{-1}$
\begin{equation}
\tilde\chi\strut_L(k,\omega)= \chi\strut_L(k)\left[1+ \frac{\chi\strut_L(k)}{\chi\strut_L(0)} M(\omega) \right]^{-1} . 
\label{eq26-1}
\end{equation}
The memory function in Eq.\ \eqref{eq23} is therefore given by the relation
\begin{equation}
M_L(k,\omega)=  -\frac{i\omega}{M(\omega)} \frac{\chi\strut_L(k)}{\chi\strut_L(0)} ,  
\end{equation}
which, with the account of Eq.\ \eqref{eq19}, reduces to the standard result,\cite{Landau8} $\epsilon(\omega)=4\pi i\sigma(\omega)/\omega$, at $k=0$.  

The static $k$-dependent susceptibility function $\chi\strut_L(k)$ in Eqs.\ \eqref{eq21} and \eqref{eq26} presents the main challenge for the modeling of ionic liquids. In order to understand the difficulties involved, one can consider the case of a non-polarizable ($\epsilon_\infty=1$)  1:1 ionic liquid carrying unit ionic charges. The static susceptibility can be related to the structure factor describing charge translations\cite{Hansen:13} 
\begin{equation}
 \chi\strut_{L}(k) = \frac{\kappa_D^2}{4\pi k^2} S^Z(k) .
 \label{eq27}
\end{equation}
Here, $\kappa_D^{-1}$ is the Debye-H{\"u}ckel length, and the structure factor of the charge density fluctuations is the sum of the component density structure factors weighted with the component charges $z_\alpha$ 
\begin{equation}
S^Z(k) = N^{-1} \sum_{\alpha,\beta}z_\alpha z_\beta \langle \rho_{\alpha,\mathbf{k}}\rho_{\beta,-\mathbf{k}}\rangle .
\label{eq28}
\end{equation}
Here, $N$ is the total number of RTIL ions, $\rho_{\alpha,\mathbf{k}}=\sum_{j}^{\alpha} e^{i\mathbf{k}\cdot\mathbf{r}_j}$ is the reciprocal-space density field of component $\alpha$, and the average $S_{\alpha\beta}(k)=(N_\alpha N_\beta)^{-1/2}\langle \rho_{\alpha,\mathbf{k}}\rho_{\beta,-\mathbf{k}}\rangle$ is the density structure factor describing  self and cross-correlations of the density fluctuations of the $\alpha$ and $\beta$ components of the mixture. The structure factor $S^Z(k)$ describes fluctuations of the charge density in a homogeneous liquid caused by translations (density fluctuations) of the particles carrying molecular charges. 

The charge structure factor $S^Z(k)$ satisfies the long wavelength limit 
\begin{equation}
(\kappa_D^2/k^2)S^Z(k)\to 1\quad k\to 0, 
\label{eq29}
\end{equation}
which guaranties $\epsilon_L(0,0)=\epsilon_s\to \infty$.\cite{Hansen:13} This condition implies strong inhibition of the charge density fluctuations at $k\to 0$ compared to the density and rotational fluctuations. Density fluctuations at $k=0$ produce a nonzero compressibility, and rotational fluctuations at $k=0$ are related to the dielectric constant. In contrast, there are no macroscopic charge density fluctuations. 

The requirement to satisfy the asymptote given by Eq.\ \eqref{eq29} makes direct modeling of polar response by dense ionic liquid a challenging task. It is often circumvented by models employing dipolar response for the longitudinal susceptibility. This is the route also adopted here. Our approach is to model $\chi\strut_L(k)$ through an effective dipolar field. The $k=0$ limit is set up through Eq.\ \eqref{eq29}, which demands $\chi\strut_L(0)=(4\pi)^{-1}$ for nonolarizable liquids and a corresponding polarizability correction according to Eq.\ \eqref{eq18} for polarizable liquids. The $k\to\infty$ limit of a dipolar function is $\chi\strut_L(\infty)=3y/(4\pi)$, where $y=(4\pi/9)\beta \rho m_\text{eff}^2$ is the standard one-particle parameter of the dielectric theories.\cite{Boettcher:73,Madden:84}  The effective squared molecular dipole $m_\text{eff}^2$ here is not well defined in the case of ionic liquids. If one accounts for ionic translations to produce one-particle dipolar fluctuations (Fig.\ \ref{fig:6}), the effective dipole should accommodate mean-square ionic displacements   
\begin{equation}
 m_\text{eff}^2 =  x_m m^2   +  \sum_\alpha x_\alpha z_\alpha^2\langle \delta \mathbf{r}_\alpha^2\rangle  
 \label{eq30}
\end{equation}
where the sum runs over the charges $z_\alpha$ and mean-square displacements $\langle \delta \mathbf{r}_\alpha^2\rangle$ of the ionic components with mole fractions $x_\alpha$. Further, $m$  
is the molecular dipole introduced above, with the molar fraction $x_m$ accounting for the fact that molecular dipole moment is often associated with one of the ions, which is the cation for $\mathrm{[bmim^+][PF_6^-]}$ ($x_m=1/2$). 

The existing evidence indicates that ionic liquids are densely packed materials, in which ions are mostly involved in glassy cage rattling\cite{DelPpolo:2004gx} on a broad range of time-scales from hundreds of picoseconds to a few nanoseconds.\cite{Arzhantsev:2007ds,Kashyap:2010db} This physical picture implies that mean-square displacements are mostly limited by the ion's cage, $\langle \delta \mathbf{r}_\alpha^2\rangle \ll  (\sigma_\alpha/2)^2$, where $\sigma_\alpha$ is the hard-sphere diameter of the component. More precise assignment is difficult to make without explicit simulations of the longitudinal structure factors of an ionic liquid.  However, the value assigned to $m_\text{eff}^2$ and, correspondingly, to $\chi\strut_L(\infty)$ does not strongly affect the results of calculations. The reason is the damping effect of the solute electrostatic potential on the result of $k$-integration in Eq.\ \eqref{eq17}. More details on the parameterization of $\chi\strut_L(k)$ can be found in the supplementary material.\cite{supmatJCP}

We consider a solute carrying the unit charge, with the radius $R_0$ placed at the distance $R$ from the electrode. The solute's electrostatic potential in reciprocal space becomes 
\begin{equation}
\phi_{0\mathbf{k}} = (4\pi e/k^2) j_0(kR_1)\left[1-e^{2i\mathbf{k}\cdot\mathbf{R}}\right] .  
\label{eq31}
\end{equation}
Here, $R_1=R_0 + \sigma/2$ is the radius of the solvent-accessible sphere, which is offset from the solute radius $R_0$ by an effective solvent radius $\sigma/2$, which needs to be determined (see below). The distance $2R$ is the separation between the ion and its image in the metal electrode and $j_0(x)$ is the spherical Bessel function of zeroth order.\cite{Abramowitz:72} 

The image effects are always present in the dielectric calculations of the electrode reorganization energy,\cite{Liu:1994ul} and result in a  linear dependence of the reorganization energy on the inverse distance $R^{-1} $ to a metal electrode. The appearance of this dependence was questioned in the past based on the perceived screening of the image forces by molecular dipoles of a polar solvent.\cite{Phelps:1990jd,Krishtalik:1991hs,Baranski:1991hy} The early calculations supporting these claims employed incorrect dipolar structure factors and were not supported by subsequent calculations.\cite{Fonseca:90,Raineri:92,DMmp:93,Perng:96} The image effects are also found to be consistent with the standard expectations in more recent simulations of electrode reactions in ionic melts.\cite{Reed:2008ez} The appearance of a substantial dependence of the thermodynamic $\lambda$ on $R$ is also supported by our present calculations (Fig.\ S1 in the supplementary material\cite{supmatJCP}). 

The ions making RTILs are typically large and have their size comparable to the size of redox pairs of the electrochemical experiment. For instance, the size of the cation in  $\mathrm{[bmim^+][PF_6^-]}$ is estimated as $6.8$ \AA\ from the van der Waals volume.\cite{Arzhantsev:2007ds} An even larger value for an effective diameter of the ionic mixture, $\sigma=7.52$ \AA, is estimated here from the compressibility of $\mathrm{[bmim^+][PF_6^-]}$ ($\sigma=7.78$ \AA\ is listed in Ref.\ \onlinecite{Kashyap:2008fe}). The approach based on compressibility typically provides a good estimate of the effective hard-sphere diameter of a liquid.\cite{BenAmotz:90,DMjpc:95} As mentioned above, fluctuations of charge density are suppressed at $k\to 0$ and do not affect the macroscopic compressibility.\cite{Hansen:13} An effective hard-sphere diameter somewhat larger than the size of individual ions can arise from strong Coulomb correlations in ionic pairs which act as single entities from the perspective of compressibility (a typical size of the ionic pair in RTIL is $\sim 8$ \AA\cite{Mezger:08}). The diameter of the ferrocene ion,\cite{Sikes:2001hp,Fietkau:2006ia} which we use here for modeling, $\sigma_0=5.2$ \AA, is smaller than the effective solvent diameter $\sigma$. The result $\sigma_0<\sigma$ makes the application of continuum solvation models, which can be justified only in the limit $\sigma_0/\sigma\gg1$, unreliable, and  nonlocal susceptibility functions, depending on the reciprocal space $k$-vector,\cite{Fried:90,Bagchi:91,DMmp:93,Perng:96} are required.

\begin{figure}
\includegraphics[clip=true,trim= 0cm 1.5cm 0cm 0cm,width=8cm]{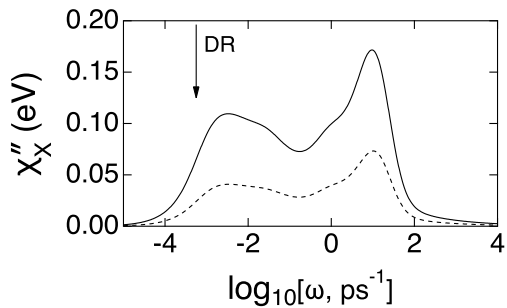}
\caption{Stokes-shift loss function for ferrocene ($R_0=2.61$ \AA, solid line) and ferrocene carboxaldehyde ($R_0=7.59$ \AA, dashed line) in $\mathrm{[bmim^+][PF_6^-]}$ RTIL: $\sigma=7.52$ \AA, $m=4.4$ D, $\epsilon_\infty =1.985$, $y=0.95$, $T=298$ K. The frequency-dependent dielectric function\cite{Stoppa:2008be} is tabulated in the supplementary material.\cite{supmatJCP} The vertical arrow indicates the lowest frequency accessible by dielectric relaxation (DR) experiment.\cite{Stoppa:2008be} 
}
\label{fig:7}
\end{figure}

We model the $\mathrm{[bmim^+][PF_6^-]}$ RTIL by using the following input parameters: the dielectric function\cite{Stoppa:2008be} $\epsilon(\omega)$, hard-sphere diameter $\sigma=7.52$ \AA, density $\rho=1.37$ g/cm$^3$, dipole moment of the cation $m=4.4$ D,\cite{Kashyap:2008fe} and the high-frequency dielectric constant $\epsilon_\infty=1.985$. The modeling of the $k\to\infty$ limit of the dipolar susceptibility mapping the charge density fluctuations of the RTIL requires specifying the dipolar density parameter $y$. The use of the effective dipole moment as given by Eq.\ \eqref{eq30} applies only to intermediate values of $k$ since $\chi_L(k)$ related to ions' translations decays to zero as $k^{-4}$, as follows from Eqs.\ \eqref{eq27} and \eqref{eq28}. In the limit $k\to\infty$ only the molecular point dipoles produce a non-zero-value of $\chi\strut_L(\infty)$. The approximation of a point molecular dipole eventually breaks down at $2\pi/k$ comparable to the intramolecular separation of charge.\cite{Raineri:99} However, that happens at in the range of wavevectors which mostly does not contribute to the solvation free energy. We therefore have chosen $\chi\strut_L(\infty)$ based on the density of molecular dipoles in the RTIL, which mostly comes from the $m=4.4$ D dipole of the cation. By adopting this approximation, we likely miss some excess polarity produced by ion translations at intermediate $k$-values and resulting in the effective dipole $m_\text{eff}$ in Eq.\ \eqref{eq30}. We want to stress that $y$ is merely an effective polarity parameter characterizing the dipolar field mapping the charge fluctuations in RTIL. As mentioned above, the limit of large $k$ does not substantially affect the Stokes-shift susceptibility determined by $k$-integration in Eq.\ \eqref{eq17}.

RTILs in contact with solid substrates form layered structures.\cite{Mezger:08} The first layer of equal-sign ions forms a plane of charge with the surface charge density determined by the size and packing of the ions at the interface. The electrostatic potential drops approximately linearly in this layer,\cite{Reed:2008ez,Vatamanu:2010we} but, in contrast to conventional solution electrolytes,\cite{BardFaulkner01} it can drop below zero (overscreening) and then approach zero with increasing distance in oscillations reflecting layered structure of the RTIL in the interface\cite{Rovere:1986iq,Bazant:2011ha,Fedorov:2014er} The oscillations are low in amplitude at the equilibrium electrode potential, but increase in amplitude when electrostatic potential is applied to the electrode.\cite{Reed:2008ez} Given this interfacial structure, it seems reasonable to assign the effective radius of the ions in the RTIL $\sigma/2$ to the thickness of the Stern layer (defined as the layer where potential drops linearly). We place a spherical solute with the diameter $\sigma_0=5.2$ \AA\, corresponding to the ferrocene cation, at the distance $R=R_0+\sigma$ from the electrode (see Fig.\ S2 in the supplementary material\cite{supmatJCP}). This placement allows us to assume that the reactant is outside the Stern layer and is in the part of the RTIL interface where oscillations of the potential decay to zero. Correspondingly, the Frumkin corrections\cite{Fedorov:2014er} for the electrode kinetics are minimized. The Frumkin corrections are generally viewed to be insignificant for RTILs.\cite{Reed:2008ez,Fedorov:2014er}

Figure \ref{fig:7} shows the Stokes-shift loss function $\chi\strut_X''(\omega)$ calculated for the ferrocene cation in $\mathrm{[bmim^+][PF_6^-]}$. Because of the screening factor by the induced dipoles in Eq.\ \eqref{eq18}, our calculations account for the effect of the solvent polarizability\cite{DMjcp1:17} often neglected in computer simulations based on non-polarizable force fields. The dependence of the loss function on the solute size is illustrated by additionally considering the cation of ferrocene carboxaldehydrate\cite{Fietkau:2006ia} with $R_0=7.59$ \AA\  (dashed line).

The loss function $\chi\strut_X''(\omega)$ is broadly distributed, in accord   with the highly stretched dielectric relaxation of RTILs.\cite{Stoppa:2008be,Sonnleitner:2015fla} Two characteristic peaks, at sub-picosecond and nanosecond range of frequencies, mirror the corresponding peaks in the dielectric loss functions. Integration of $\chi\strut_X''(\omega)$ in Eq.\ \eqref{eq2-2} yields the nonergodic reorganization energy $\lambda(k_\text{el})$ (Fig.\ \ref{fig:1}). It slowly increases, with lowering $k_\text{el}$, over a very broad range of rate values to a saturation magnitude resulting from the cutoff of the frequency domain in the dielectric experiment.\cite{Stoppa:2008be} Since low-frequency processes, not resolved by the dielectric experiment, are not included in our model of the dielectric constant, the dielectric function remains constant at low frequencies, thus producing a nearly constant value of the reorganization energy.  Its value is below the thermodynamic reorganization energy,\cite{Nikitina:2014ei,comLambda} $\lambda(0)\simeq 0.68$ eV, estimated here by putting $\epsilon(\omega)=\epsilon_s\to \infty$ in the susceptibility functions in Eqs.\ \eqref{eq26} and \eqref{eq26-1}.

\begin{figure}
\includegraphics[clip=true,trim= 0cm 1.5cm 0cm 0cm,width=8cm]{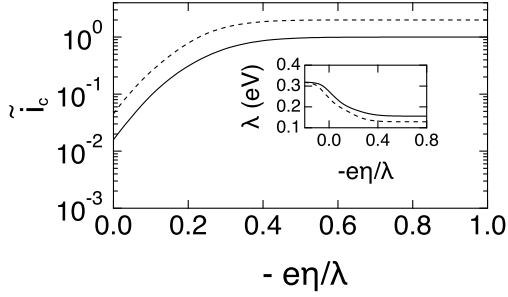}
\caption{Scaled cathodic current $\tilde i_c=i_c\tau_s/(A\Gamma_t)$, $\tau_s=1$ ps vs the electrode overpotential $\eta$ scaled with the thermodynamic reorganization energy $\lambda=\lambda(0)=0.68$ eV for cathodic reaction of the ferrocene cation in the $\mathrm{[bmim^+][PF_6^-]}$ RTIL. The inset shows the reduction of the nonergodic reorganization energy $\lambda(\eta)$ due to the increase in the reaction rate. The parameters of the solvent are the same as in Figs.\ \ref{fig:1} and \ref{fig:7}; $\kappa_s=1$ (solid lines) and $\kappa=2$ (dashed lines) is adopted in the rate calculations.    
}
\label{fig:8}
\end{figure}
   
A shallow dependence of the reorganization energy on the rate constant, much different from a sharp change found for the Debye process (Fig.\ \ref{fig:2}), is the result of the stretched exponential dynamics specific to the RTILs. A very similar slow variation of the reorganization energy with the changing observation window was found in extensive computer simulations of proteins extending to the microsecond length of trajectories.\cite{DMjcp1:15} The underlying phenomenology is likely common to both cases given a highly stretched dielectric relaxation of proteins.\cite{Hong:2011qf,Nakanishi:2014gy} Consistent with the slow alteration in the reorganization energy, there are no discontinuities in the cathodic current curves (Fig.\ \ref{fig:8}). The function $\lambda(\eta)$ is shown in the inset in Fig.\ \ref{fig:8}. The reorganization energy is still changing with the overpotential due to nonergotic restrictions on the available phase space, but no discontinuities are seen in this case, in contrast to the Debye relaxation scenario shown in Fig.\ \ref{fig:2}.

\begin{figure}
\includegraphics[clip=true,trim= 0cm 1.5cm 0cm 0cm,width=8cm]{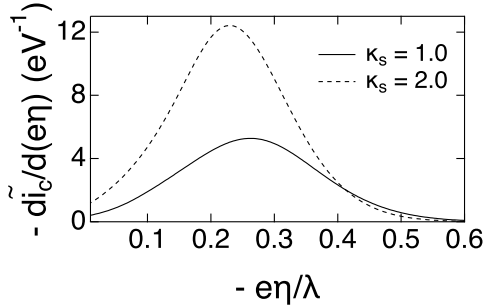}
\caption{$-d\tilde i_c/d(e\eta)$ calculated for ferrocene in  $\mathrm{[bmim^+][PF_6^-]}$; $\kappa_s=1$ (solid line) and $\kappa_s=2.0$ (dashed line).  The rest of the parameters are the same as in Figs.\ \ref{fig:7} and \ref{fig:8}.  }
\label{fig:9}
\end{figure}

The dependence of the reorganization energy on overpotential should cause a generally non-parabolic form for the activation barrier $\Delta G^\dag(\eta)$, as was already shown for a solvent with Debye relaxation in Fig.\ \ref{fig:4}. Since the dependence $\lambda(\eta)$ is significantly smoothed out in the case of the RTILs, one does not observe sharp features in the derivative of the current with respect to the overpotential and, instead, a smooth curve follows (Fig.\ \ref{fig:9}). It is shallower on its left wing, reflecting a higher reorganization energy at less negative overpotentials (Fig.\ \ref{fig:8} inset).  

The thermodynamic reorganization energy shown in Fig.\ \ref{fig:1} is close in magnitude to what one anticipates for molecular polar solvents.\cite{Lynden-Bell:2007pd,LyndenBell:2007ff,Shim:2009jf} However, since the relaxation of RTILs is much slower, fast electrode reactions in RTILs have puzzled many researchers.\cite{Nikitina:2014ei,Mladenova:2016gu}  Our resolution of this puzzle is in terms of nonergodic effects on the solvent reorganization energy: the nonergodic character of the electrode reaction, due to complex dynamics of RTILs, leads to a significantly lower reorganization energy (Fig.\ \ref{fig:1}) and a correspondingly lower activation barrier.  We note that our results generally apply to nuclear solvation produced by RTILs. A corresponding reduction of both the reorganization energy and of the driving force is expected for homogeneous electron-transfer reactions between the reactants dissolved in RTILs.

\section{Discussion and conclusions}
The model presented here deals with the consequences of reaction nonergodicity for electrode reactions in media with sufficiently slow relaxation such that the relaxation time and the reaction time are comparable in magnitude. This condition leads to breaking of ergodicity for the statistics of configurations in the reactant and product wells (Fig.\ \ref{fig:0}). Ergodicity breaking, and the related difficulties in formulating the Gibbs distribution,\cite{Palmer:82} is a general phenomenon most commonly encountered in phase transitions and relaxation of glassy materials.\cite{Crisanti:00} It reflects the inability of a statistical ensemble to completely sample all its allowed phase space within the time of observation. The problem is circumvented by defining Gibbs-weighted averages on a restricted sub-space of the entire phase space of the system.\cite{Palmer:82} In other words, the statistical averages are calculated with the Boltzmann-Gibbs weights within a subspace $\Gamma$
\begin{equation}
\langle \dots \rangle \propto \int_\Gamma d\Gamma \dots e^{-\beta H} .  
\label{eq32}
\end{equation}
This formalism assumes that only configurational fluctuations can be slow and momentum transfer is always fast, making the kinetic temperature ($\beta=1/(k_\text{B}T)$ in Eq.\ \eqref{eq32}) well-defined.

The definition of the restricted phase space ($\Gamma$ in Eq.\ \eqref{eq32}) is less straightforward for activated transitions, when the activation barrier can be altered by either changing the thermodynamic state or some external conditions. This difficulty becomes particularly severe for chemical reactions. In this case, the rate of the reaction itself defines the observation window, which therefore has to be adjusted self-continuously with the account for the dependence of the activation barrier on the rate (Eq.\ \eqref{eq2}). The restricted ensemble becomes dynamically restricted, which is achieved by specifying the phase space region $\Gamma$ in Eq.\ \eqref{eq32}
\begin{equation}
\langle \dots \rangle \propto \int_{\tau_s k < 1} d\Gamma \dots e^{-\beta H} .  
\label{eq33}
\end{equation}

The result of this definition of statistical averages is the appearance of the dependence on the rate constant in the free energy assigned to reactants and products. When applied to the rate of activated transitions, this notion leads to the formulation of the nonergodic chemical kinetics.\cite{DMjcp2:05,DMjpcm:15}  Precursors of this concept can be found in the isomerization theory of van der Zwan and Hynes\cite{Zwan:85} and in the Sumi-Marcus theory of the dynamic solvent effect on electron transfer.\cite{Sumi:86} In both cases, solutions for the two-dimensional barrier crossing problem allow the dynamics to drive the reaction along a path deviating from the lowest barrier height. Nonergodic kinetics replaces the dynamic solution with an ensemble perspective (similar to the replacement of the Newtonian dynamics with the Gibbs canonical ensemble).\cite{DMjcp1:09} The advantage of this approach is that it is not limited to a small number of dynamical coordinates and can be applied to systems with many stochastic modes coupled to the reaction coordinate and producing complex dynamics not allowed by simple dynamic models.     

Electrode reactions are a particularly important class of activated proccesses  because of the freedom to alter the reaction free energy and the activation barrier  through the electrode overpotential. One can therefore drive the reaction to the nonergodic regime by merely sweeping the electrode potential. Here, we formulated a theory of electrode reactions in slowly relaxing media and have shown that switching to nonergodic kinetics is indeed possible by adjusting the electrode overpotential.             

The transition from the kinetics based on the free energy of activation following from the Gibbs ensemble (thermodynamics) to the nonergodic regime requiring account of the relaxation times is rather abrupt with the sweep of the electrode potential when the relaxation of the medium is characterized by a single Debye process. The transition occurs when the equality condition (Eqs.\ \eqref{eq0}  and \eqref{eq8}) between the rate $k_\text{el}(\eta)$ and the Debye relaxation time $\tau_s$ is met. Allowing a broad distribution of relaxation times, which is the situation realized for RTILs, broadens the transition and leads to a relatively shallow dependence of the activation parameters (reorganization energy here) on the external tuning of the activation barrier. However, the dependence of the reorganization energy on the electrode overpotential is preserved in this case as well (Fig.\ \ref{fig:8} inset). It leads to a generally non-parabolic dependence of the activation barrier on the electrode overpotential. The nonergodic  reorganization energy of electron transfer in RTILs depends on the observation window specified by the reaction rate. It slowly changes when the reaction rate is altered over several orders of magnitude (Fig.\ \ref{fig:1}), similarly to electron transfer in proteins.\cite{DMjcp1:15} The phenomenology found here seems to be quite general and there are significant reasons to believe that this picture should be common to a number of materials with stretched exponential dynamics.

\section*{Supplementary material}
See supplementary material for the parametrization of the nonlocal susceptibilities of RTILs. 

\acknowledgements 
This research was supported by the Office of Basic Energy Sciences, Division of Chemical Sciences, Geosciences, and Energy Biosciences, Department of Energy (DE-SC0015641).


\begin{thebibliography}{125}%
\makeatletter
\providecommand \@ifxundefined [1]{%
 \@ifx{#1\undefined}
}%
\providecommand \@ifnum [1]{%
 \ifnum #1\expandafter \@firstoftwo
 \else \expandafter \@secondoftwo
 \fi
}%
\providecommand \@ifx [1]{%
 \ifx #1\expandafter \@firstoftwo
 \else \expandafter \@secondoftwo
 \fi
}%
\providecommand \natexlab [1]{#1}%
\providecommand \enquote  [1]{``#1''}%
\providecommand \bibnamefont  [1]{#1}%
\providecommand \bibfnamefont [1]{#1}%
\providecommand \citenamefont [1]{#1}%
\providecommand \href@noop [0]{\@secondoftwo}%
\providecommand \href [0]{\begingroup \@sanitize@url \@href}%
\providecommand \@href[1]{\@@startlink{#1}\@@href}%
\providecommand \@@href[1]{\endgroup#1\@@endlink}%
\providecommand \@sanitize@url [0]{\catcode `\\12\catcode `\$12\catcode
  `\&12\catcode `\#12\catcode `\^12\catcode `\_12\catcode `\%12\relax}%
\providecommand \@@startlink[1]{}%
\providecommand \@@endlink[0]{}%
\providecommand \url  [0]{\begingroup\@sanitize@url \@url }%
\providecommand \@url [1]{\endgroup\@href {#1}{\urlprefix }}%
\providecommand \urlprefix  [0]{URL }%
\providecommand \Eprint [0]{\href }%
\providecommand \doibase [0]{http://dx.doi.org/}%
\providecommand \selectlanguage [0]{\@gobble}%
\providecommand \bibinfo  [0]{\@secondoftwo}%
\providecommand \bibfield  [0]{\@secondoftwo}%
\providecommand \translation [1]{[#1]}%
\providecommand \BibitemOpen [0]{}%
\providecommand \bibitemStop [0]{}%
\providecommand \bibitemNoStop [0]{.\EOS\space}%
\providecommand \EOS [0]{\spacefactor3000\relax}%
\providecommand \BibitemShut  [1]{\csname bibitem#1\endcsname}%
\let\auto@bib@innerbib\@empty
\bibitem [{\citenamefont {Marcus}(1965)}]{Marcus:65}%
  \BibitemOpen
  \bibfield  {author} {\bibinfo {author} {\bibfnamefont {R.~A.}\ \bibnamefont
  {Marcus}},\ }\href@noop {} {\bibfield  {journal} {\bibinfo  {journal} {J.\
  Chem.\ Phys.}\ }\textbf {\bibinfo {volume} {43}},\ \bibinfo {pages} {679}
  (\bibinfo {year} {1965})}\BibitemShut {NoStop}%
\bibitem [{\citenamefont {Hush}(1968)}]{Hush:1968rr}%
  \BibitemOpen
  \bibfield  {author} {\bibinfo {author} {\bibfnamefont {N.~S.}\ \bibnamefont
  {Hush}},\ }\href {\doibase http://dx.doi.org/10.1016/0013-4686(68)80032-5}
  {\bibfield  {journal} {\bibinfo  {journal} {Electrochim. Acta}\ }\textbf
  {\bibinfo {volume} {13}},\ \bibinfo {pages} {1005} (\bibinfo {year}
  {1968})}\BibitemShut {NoStop}%
\bibitem [{\citenamefont {Hush}(1999)}]{Hush:99}%
  \BibitemOpen
  \bibfield  {author} {\bibinfo {author} {\bibfnamefont {N.~S.}\ \bibnamefont
  {Hush}},\ }\href@noop {} {\bibfield  {journal} {\bibinfo  {journal} {J.\
  Electroanal.\ Chem.}\ }\textbf {\bibinfo {volume} {470}},\ \bibinfo {pages}
  {170} (\bibinfo {year} {1999})}\BibitemShut {NoStop}%
\bibitem [{\citenamefont {Levich}(1965)}]{Levich:1965wp}%
  \BibitemOpen
  \bibfield  {author} {\bibinfo {author} {\bibfnamefont {V.~G.}\ \bibnamefont
  {Levich}},\ }in\ \href@noop {} {\emph {\bibinfo {booktitle} {Advances in
  Electrochemistry and Electrochemical Engineering}}},\ Vol.~\bibinfo {volume}
  {4},\ \bibinfo {editor} {edited by\ \bibinfo {editor} {\bibfnamefont
  {P.}~\bibnamefont {Delahay}}}\ (\bibinfo  {publisher} {Interscience},\
  \bibinfo {address} {New York},\ \bibinfo {year} {1965})\ pp.\ \bibinfo
  {pages} {1--124}\BibitemShut {NoStop}%
\bibitem [{\citenamefont {Chidsey}(1991)}]{Chidsey:91}%
  \BibitemOpen
  \bibfield  {author} {\bibinfo {author} {\bibfnamefont {C.~E.~D.}\
  \bibnamefont {Chidsey}},\ }\href@noop {} {\bibfield  {journal} {\bibinfo
  {journal} {Science}\ }\textbf {\bibinfo {volume} {251}},\ \bibinfo {pages}
  {919} (\bibinfo {year} {1991})}\BibitemShut {NoStop}%
\bibitem [{\citenamefont {Finklea}\ \emph {et~al.}(2001)\citenamefont
  {Finklea}, \citenamefont {Yoon}, \citenamefont {Chamberlain}, \citenamefont
  {Allen},\ and\ \citenamefont {Haddox}}]{Finklea:2001ve}%
  \BibitemOpen
  \bibfield  {author} {\bibinfo {author} {\bibfnamefont {H.~O.}\ \bibnamefont
  {Finklea}}, \bibinfo {author} {\bibfnamefont {K.}~\bibnamefont {Yoon}},
  \bibinfo {author} {\bibfnamefont {E.}~\bibnamefont {Chamberlain}}, \bibinfo
  {author} {\bibfnamefont {J.}~\bibnamefont {Allen}}, \ and\ \bibinfo {author}
  {\bibfnamefont {R.}~\bibnamefont {Haddox}},\ }\href {\doibase
  10.1021/jp0041510} {\bibfield  {journal} {\bibinfo  {journal} {J. Phys. Chem.
  B}\ }\textbf {\bibinfo {volume} {105}},\ \bibinfo {pages} {3088} (\bibinfo
  {year} {2001})}\BibitemShut {NoStop}%
\bibitem [{\citenamefont {Sav{\'e}ant}(2002)}]{Saveant:2002em}%
  \BibitemOpen
  \bibfield  {author} {\bibinfo {author} {\bibfnamefont {J.-M.}\ \bibnamefont
  {Sav{\'e}ant}},\ }\href@noop {} {\bibfield  {journal} {\bibinfo  {journal}
  {J. Phys. Chem. B}\ }\textbf {\bibinfo {volume} {106}},\ \bibinfo {pages}
  {9387} (\bibinfo {year} {2002})}\BibitemShut {NoStop}%
\bibitem [{\citenamefont {Gosavi}\ and\ \citenamefont
  {Marcus}(2000)}]{Gosavi:00}%
  \BibitemOpen
  \bibfield  {author} {\bibinfo {author} {\bibfnamefont {S.}~\bibnamefont
  {Gosavi}}\ and\ \bibinfo {author} {\bibfnamefont {R.~A.}\ \bibnamefont
  {Marcus}},\ }\href@noop {} {\bibfield  {journal} {\bibinfo  {journal} {J.
  Phys. Chem. B}\ }\textbf {\bibinfo {volume} {104}},\ \bibinfo {pages} {2067}
  (\bibinfo {year} {2000})}\BibitemShut {NoStop}%
\bibitem [{\citenamefont {Gosavi}, \citenamefont {Gao},\ and\ \citenamefont
  {Marcus}(2001)}]{Gosavi:2001jh}%
  \BibitemOpen
  \bibfield  {author} {\bibinfo {author} {\bibfnamefont {S.}~\bibnamefont
  {Gosavi}}, \bibinfo {author} {\bibfnamefont {Y.~Q.}\ \bibnamefont {Gao}}, \
  and\ \bibinfo {author} {\bibfnamefont {R.~A.}\ \bibnamefont {Marcus}},\
  }\href@noop {} {\bibfield  {journal} {\bibinfo  {journal} {J. Electroanal.
  Chem.}\ }\textbf {\bibinfo {volume} {500}},\ \bibinfo {pages} {71} (\bibinfo
  {year} {2001})}\BibitemShut {NoStop}%
\bibitem [{\citenamefont {Migliore}\ and\ \citenamefont
  {Nitzan}(2012)}]{Migliore:2012kx}%
  \BibitemOpen
  \bibfield  {author} {\bibinfo {author} {\bibfnamefont {A.}~\bibnamefont
  {Migliore}}\ and\ \bibinfo {author} {\bibfnamefont {A.}~\bibnamefont
  {Nitzan}},\ }\href {\doibase
  http://dx.doi.org/10.1016/j.jelechem.2012.02.026} {\bibfield  {journal}
  {\bibinfo  {journal} {J. Electroanal. Chem.}\ }\textbf {\bibinfo {volume}
  {671}},\ \bibinfo {pages} {99} (\bibinfo {year} {2012})}\BibitemShut
  {NoStop}%
\bibitem [{\citenamefont {Zusman}(1980)}]{Zusman:80}%
  \BibitemOpen
  \bibfield  {author} {\bibinfo {author} {\bibfnamefont {L.~D.}\ \bibnamefont
  {Zusman}},\ }\href@noop {} {\bibfield  {journal} {\bibinfo  {journal} {Chem.
  Phys.}\ }\textbf {\bibinfo {volume} {49}},\ \bibinfo {pages} {295} (\bibinfo
  {year} {1980})}\BibitemShut {NoStop}%
\bibitem [{\citenamefont {Sumi}\ and\ \citenamefont {Marcus}(1986)}]{Sumi:86}%
  \BibitemOpen
  \bibfield  {author} {\bibinfo {author} {\bibfnamefont {H.}~\bibnamefont
  {Sumi}}\ and\ \bibinfo {author} {\bibfnamefont {R.~A.}\ \bibnamefont
  {Marcus}},\ }\href@noop {} {\bibfield  {journal} {\bibinfo  {journal} {J.
  Chem. Phys}\ }\textbf {\bibinfo {volume} {84}},\ \bibinfo {pages} {4894}
  (\bibinfo {year} {1986})}\BibitemShut {NoStop}%
\bibitem [{\citenamefont {Rips}\ and\ \citenamefont
  {Jortner}(1987)}]{Rips:1987qf}%
  \BibitemOpen
  \bibfield  {author} {\bibinfo {author} {\bibfnamefont {I.}~\bibnamefont
  {Rips}}\ and\ \bibinfo {author} {\bibfnamefont {J.}~\bibnamefont {Jortner}},\
  }\href@noop {} {\bibfield  {journal} {\bibinfo  {journal} {J. Chem. Phys.}\
  }\textbf {\bibinfo {volume} {87}},\ \bibinfo {pages} {2090} (\bibinfo {year}
  {1987})}\BibitemShut {NoStop}%
\bibitem [{\citenamefont {Yan}, \citenamefont {Sparpaglione},\ and\
  \citenamefont {Mukamel}(1988)}]{Yan:1988mz}%
  \BibitemOpen
  \bibfield  {author} {\bibinfo {author} {\bibfnamefont {Y.~J.}\ \bibnamefont
  {Yan}}, \bibinfo {author} {\bibfnamefont {M.}~\bibnamefont {Sparpaglione}}, \
  and\ \bibinfo {author} {\bibfnamefont {S.}~\bibnamefont {Mukamel}},\ }\href
  {\doibase 10.1021/j100328a010} {\bibfield  {journal} {\bibinfo  {journal} {J.
  Phys. Chem.}\ }\textbf {\bibinfo {volume} {92}},\ \bibinfo {pages} {4842}
  (\bibinfo {year} {1988})}\BibitemShut {NoStop}%
\bibitem [{\citenamefont {Smith}, \citenamefont {Staib},\ and\ \citenamefont
  {Hynes}(1993)}]{Smith:1993dx}%
  \BibitemOpen
  \bibfield  {author} {\bibinfo {author} {\bibfnamefont {B.~B.}\ \bibnamefont
  {Smith}}, \bibinfo {author} {\bibfnamefont {A.}~\bibnamefont {Staib}}, \ and\
  \bibinfo {author} {\bibfnamefont {J.~T.}\ \bibnamefont {Hynes}},\ }\href@noop
  {} {\bibfield  {journal} {\bibinfo  {journal} {Chem. Phys.}\ }\textbf
  {\bibinfo {volume} {176}},\ \bibinfo {pages} {521} (\bibinfo {year}
  {1993})}\BibitemShut {NoStop}%
\bibitem [{\citenamefont {Morgan}\ and\ \citenamefont
  {Wolynes}(1987)}]{Morgan:1987bm}%
  \BibitemOpen
  \bibfield  {author} {\bibinfo {author} {\bibfnamefont {J.~D.}\ \bibnamefont
  {Morgan}}\ and\ \bibinfo {author} {\bibfnamefont {P.~G.}\ \bibnamefont
  {Wolynes}},\ }\href@noop {} {\bibfield  {journal} {\bibinfo  {journal} {J.
  Phys. Chem.}\ }\textbf {\bibinfo {volume} {91}},\ \bibinfo {pages} {874}
  (\bibinfo {year} {1987})}\BibitemShut {NoStop}%
\bibitem [{\citenamefont {Chakravarti}\ and\ \citenamefont
  {Sebastian}(1992)}]{Chakravarti:1992ee}%
  \BibitemOpen
  \bibfield  {author} {\bibinfo {author} {\bibfnamefont {N.}~\bibnamefont
  {Chakravarti}}\ and\ \bibinfo {author} {\bibfnamefont {K.~L.}\ \bibnamefont
  {Sebastian}},\ }\href@noop {} {\bibfield  {journal} {\bibinfo  {journal}
  {Chem. Phys. Lett.}\ }\textbf {\bibinfo {volume} {193}},\ \bibinfo {pages}
  {456} (\bibinfo {year} {1992})}\BibitemShut {NoStop}%
\bibitem [{\citenamefont {Matyushov}(1994)}]{DMjec:94}%
  \BibitemOpen
  \bibfield  {author} {\bibinfo {author} {\bibfnamefont {D.}~\bibnamefont
  {Matyushov}},\ }\href@noop {} {\bibfield  {journal} {\bibinfo  {journal} {J.
  Electroanal. Chem.}\ }\textbf {\bibinfo {volume} {367}},\ \bibinfo {pages}
  {1} (\bibinfo {year} {1994})}\BibitemShut {NoStop}%
\bibitem [{\citenamefont {Schmickler}(1986)}]{Schmickler:86}%
  \BibitemOpen
  \bibfield  {author} {\bibinfo {author} {\bibfnamefont {W.}~\bibnamefont
  {Schmickler}},\ }\href@noop {} {\bibfield  {journal} {\bibinfo  {journal} {J.
  Electroanal. Chem.}\ }\textbf {\bibinfo {volume} {204}},\ \bibinfo {pages}
  {31} (\bibinfo {year} {1986})}\BibitemShut {NoStop}%
\bibitem [{\citenamefont {Gorodyskii}, \citenamefont {Karasevskii},\ and\
  \citenamefont {Matyushov}(1991)}]{DMjec:91}%
  \BibitemOpen
  \bibfield  {author} {\bibinfo {author} {\bibfnamefont {A.~V.}\ \bibnamefont
  {Gorodyskii}}, \bibinfo {author} {\bibfnamefont {A.~I.}\ \bibnamefont
  {Karasevskii}}, \ and\ \bibinfo {author} {\bibfnamefont {D.~V.}\ \bibnamefont
  {Matyushov}},\ }\href@noop {} {\bibfield  {journal} {\bibinfo  {journal} {J.
  Electroanal. Chem.}\ }\textbf {\bibinfo {volume} {315}},\ \bibinfo {pages}
  {9} (\bibinfo {year} {1991})}\BibitemShut {NoStop}%
\bibitem [{\citenamefont {Schmickler}(1996)}]{Schmickler:96}%
  \BibitemOpen
  \bibfield  {author} {\bibinfo {author} {\bibfnamefont {W.}~\bibnamefont
  {Schmickler}},\ }\href@noop {} {\emph {\bibinfo {title} {Interfacial
  Electrochemistry}}}\ (\bibinfo  {publisher} {Oxford University Press},\
  \bibinfo {address} {New York},\ \bibinfo {year} {1996})\BibitemShut {NoStop}%
\bibitem [{\citenamefont {Matyushov}(2009{\natexlab{a}})}]{DMjcp2:09}%
  \BibitemOpen
  \bibfield  {author} {\bibinfo {author} {\bibfnamefont {D.~V.}\ \bibnamefont
  {Matyushov}},\ }\href@noop {} {\bibfield  {journal} {\bibinfo  {journal} {J.\
  Chem.\ Phys.}\ }\textbf {\bibinfo {volume} {130}},\ \bibinfo {pages} {234704}
  (\bibinfo {year} {2009}{\natexlab{a}})}\BibitemShut {NoStop}%
\bibitem [{\citenamefont {Kramers}(1940)}]{Kramers:1940wm}%
  \BibitemOpen
  \bibfield  {author} {\bibinfo {author} {\bibfnamefont {H.}~\bibnamefont
  {Kramers}},\ }\href@noop {} {\bibfield  {journal} {\bibinfo  {journal}
  {Physica}\ }\textbf {\bibinfo {volume} {7}},\ \bibinfo {pages} {284}
  (\bibinfo {year} {1940})}\BibitemShut {NoStop}%
\bibitem [{\citenamefont {Northrup}\ and\ \citenamefont
  {Hynes}(1980)}]{Northrup:1980hw}%
  \BibitemOpen
  \bibfield  {author} {\bibinfo {author} {\bibfnamefont {S.~H.}\ \bibnamefont
  {Northrup}}\ and\ \bibinfo {author} {\bibfnamefont {J.~T.}\ \bibnamefont
  {Hynes}},\ }\href@noop {} {\bibfield  {journal} {\bibinfo  {journal} {J.
  Chem. Phys.}\ }\textbf {\bibinfo {volume} {73}},\ \bibinfo {pages} {2700}
  (\bibinfo {year} {1980})}\BibitemShut {NoStop}%
\bibitem [{\citenamefont {Hansen}\ and\ \citenamefont
  {McDonald}(2013)}]{Hansen:13}%
  \BibitemOpen
  \bibfield  {author} {\bibinfo {author} {\bibfnamefont {J.-P.}\ \bibnamefont
  {Hansen}}\ and\ \bibinfo {author} {\bibfnamefont {I.~R.}\ \bibnamefont
  {McDonald}},\ }\href@noop {} {\emph {\bibinfo {title} {Theory of simple
  liquids}}},\ \bibinfo {edition} {4th}\ ed.\ (\bibinfo  {publisher} {Academic
  Press},\ \bibinfo {address} {Amsterdam},\ \bibinfo {year} {2013})\BibitemShut
  {NoStop}%
\bibitem [{\citenamefont {Matyushov}(2007)}]{DMacc:07}%
  \BibitemOpen
  \bibfield  {author} {\bibinfo {author} {\bibfnamefont {D.~V.}\ \bibnamefont
  {Matyushov}},\ }\href@noop {} {\bibfield  {journal} {\bibinfo  {journal}
  {Acc. Chem. Res.}\ }\textbf {\bibinfo {volume} {40}},\ \bibinfo {pages} {294}
  (\bibinfo {year} {2007})}\BibitemShut {NoStop}%
\bibitem [{\citenamefont {Matyushov}(2009{\natexlab{b}})}]{DMjcp1:09}%
  \BibitemOpen
  \bibfield  {author} {\bibinfo {author} {\bibfnamefont {D.~V.}\ \bibnamefont
  {Matyushov}},\ }\href@noop {} {\bibfield  {journal} {\bibinfo  {journal} {J.\
  Chem.\ Phys.}\ }\textbf {\bibinfo {volume} {130}},\ \bibinfo {pages} {164522}
  (\bibinfo {year} {2009}{\natexlab{b}})}\BibitemShut {NoStop}%
\bibitem [{\citenamefont {LeBard}\ and\ \citenamefont
  {Matyushov}(2008)}]{DMjpcb1:08}%
  \BibitemOpen
  \bibfield  {author} {\bibinfo {author} {\bibfnamefont {D.~N.}\ \bibnamefont
  {LeBard}}\ and\ \bibinfo {author} {\bibfnamefont {D.~V.}\ \bibnamefont
  {Matyushov}},\ }\href@noop {} {\bibfield  {journal} {\bibinfo  {journal} {J.
  Phys. Chem. B}\ }\textbf {\bibinfo {volume} {112}},\ \bibinfo {pages} {5218}
  (\bibinfo {year} {2008})}\BibitemShut {NoStop}%
\bibitem [{\citenamefont {LeBard}, \citenamefont {Kapko},\ and\ \citenamefont
  {Matyushov}(2008)}]{DMjpcb2:08}%
  \BibitemOpen
  \bibfield  {author} {\bibinfo {author} {\bibfnamefont {D.~N.}\ \bibnamefont
  {LeBard}}, \bibinfo {author} {\bibfnamefont {V.}~\bibnamefont {Kapko}}, \
  and\ \bibinfo {author} {\bibfnamefont {D.~V.}\ \bibnamefont {Matyushov}},\
  }\href@noop {} {\bibfield  {journal} {\bibinfo  {journal} {J. Phys. Chem. B}\
  }\textbf {\bibinfo {volume} {112}},\ \bibinfo {pages} {10322} (\bibinfo
  {year} {2008})}\BibitemShut {NoStop}%
\bibitem [{\citenamefont {Tafel}(1905)}]{Tafel:05}%
  \BibitemOpen
  \bibfield  {author} {\bibinfo {author} {\bibfnamefont {J.}~\bibnamefont
  {Tafel}},\ }\href@noop {} {\bibfield  {journal} {\bibinfo  {journal} {Z.\
  Phys.\ Chem.}\ }\textbf {\bibinfo {volume} {50}},\ \bibinfo {pages} {641}
  (\bibinfo {year} {1905})}\BibitemShut {NoStop}%
\bibitem [{\citenamefont {Bard}\ and\ \citenamefont
  {Faulkner}(2001)}]{BardFaulkner01}%
  \BibitemOpen
  \bibfield  {author} {\bibinfo {author} {\bibfnamefont {A.~J.}\ \bibnamefont
  {Bard}}\ and\ \bibinfo {author} {\bibfnamefont {L.~R.}\ \bibnamefont
  {Faulkner}},\ }\href@noop {} {\emph {\bibinfo {title} {Electrochemical
  Methods. Fundamentals and Applications}}},\ \bibinfo {edition} {2nd}\ ed.\
  (\bibinfo  {publisher} {Wiley},\ \bibinfo {address} {New York},\ \bibinfo
  {year} {2001})\BibitemShut {NoStop}%
\bibitem [{\citenamefont {Eyring}, \citenamefont {Lin},\ and\ \citenamefont
  {Lin.}(1980)}]{Eyring:80}%
  \BibitemOpen
  \bibfield  {author} {\bibinfo {author} {\bibfnamefont {H.}~\bibnamefont
  {Eyring}}, \bibinfo {author} {\bibfnamefont {S.~H.}\ \bibnamefont {Lin}}, \
  and\ \bibinfo {author} {\bibfnamefont {S.~M.}\ \bibnamefont {Lin.}},\
  }\href@noop {} {\emph {\bibinfo {title} {Basic {C}hemical {K}inetics}}}\
  (\bibinfo  {publisher} {Wiley-Interscience},\ \bibinfo {address} {New York},\
  \bibinfo {year} {1980})\BibitemShut {NoStop}%
\bibitem [{\citenamefont
  {B{{\"o}}ttcher}(1973{\natexlab{a}})}]{BoettcherII:73}%
  \BibitemOpen
  \bibfield  {author} {\bibinfo {author} {\bibfnamefont {C.~J.~F.}\
  \bibnamefont {B{{\"o}}ttcher}},\ }\href@noop {} {\emph {\bibinfo {title}
  {Theory of electric polarization}}},\ Vol.~\bibinfo {volume} {2}\ (\bibinfo
  {publisher} {Elsevier},\ \bibinfo {year} {1973})\BibitemShut {NoStop}%
\bibitem [{\citenamefont {Ediger}, \citenamefont {Angell},\ and\ \citenamefont
  {Nagel}(1996)}]{Ediger:96}%
  \BibitemOpen
  \bibfield  {author} {\bibinfo {author} {\bibfnamefont {M.~D.}\ \bibnamefont
  {Ediger}}, \bibinfo {author} {\bibfnamefont {C.~A.}\ \bibnamefont {Angell}},
  \ and\ \bibinfo {author} {\bibfnamefont {S.~R.}\ \bibnamefont {Nagel}},\
  }\href@noop {} {\bibfield  {journal} {\bibinfo  {journal} {J.\ Phys.\ Chem.}\
  }\textbf {\bibinfo {volume} {100}},\ \bibinfo {pages} {13200} (\bibinfo
  {year} {1996})}\BibitemShut {NoStop}%
\bibitem [{\citenamefont {Fietkau}\ \emph {et~al.}(2006)\citenamefont
  {Fietkau}, \citenamefont {Clegg}, \citenamefont {Evans}, \citenamefont
  {Villagr~n}, \citenamefont {Hardacre},\ and\ \citenamefont
  {Compton}}]{Fietkau:2006ia}%
  \BibitemOpen
  \bibfield  {author} {\bibinfo {author} {\bibfnamefont {N.}~\bibnamefont
  {Fietkau}}, \bibinfo {author} {\bibfnamefont {A.~D.}\ \bibnamefont {Clegg}},
  \bibinfo {author} {\bibfnamefont {R.~G.}\ \bibnamefont {Evans}}, \bibinfo
  {author} {\bibfnamefont {C.}~\bibnamefont {Villagr~n}}, \bibinfo {author}
  {\bibfnamefont {C.}~\bibnamefont {Hardacre}}, \ and\ \bibinfo {author}
  {\bibfnamefont {R.~G.}\ \bibnamefont {Compton}},\ }\href@noop {} {\bibfield
  {journal} {\bibinfo  {journal} {ChemPhysChem}\ }\textbf {\bibinfo {volume}
  {7}},\ \bibinfo {pages} {1041} (\bibinfo {year} {2006})}\BibitemShut
  {NoStop}%
\bibitem [{\citenamefont {Khoshtariya}, \citenamefont {Dolidze},\ and\
  \citenamefont {van Eldik}(2009)}]{Khoshtariya:2009mi}%
  \BibitemOpen
  \bibfield  {author} {\bibinfo {author} {\bibfnamefont {D.~E.}\ \bibnamefont
  {Khoshtariya}}, \bibinfo {author} {\bibfnamefont {T.~D.}\ \bibnamefont
  {Dolidze}}, \ and\ \bibinfo {author} {\bibfnamefont {R.}~\bibnamefont {van
  Eldik}},\ }\href {http://dx.doi.org/10.1002/chem.200802450} {\bibfield
  {journal} {\bibinfo  {journal} {Chemistry --A European Journal}\ }\textbf
  {\bibinfo {volume} {15}},\ \bibinfo {pages} {5254} (\bibinfo {year}
  {2009})}\BibitemShut {NoStop}%
\bibitem [{\citenamefont {Fawcett}, \citenamefont {Ga{\'a}l},\ and\
  \citenamefont {Misicak}(2011)}]{Fawcett:2011fp}%
  \BibitemOpen
  \bibfield  {author} {\bibinfo {author} {\bibfnamefont {W.~R.}\ \bibnamefont
  {Fawcett}}, \bibinfo {author} {\bibfnamefont {A.}~\bibnamefont {Ga{\'a}l}}, \
  and\ \bibinfo {author} {\bibfnamefont {D.}~\bibnamefont {Misicak}},\
  }\href@noop {} {\bibfield  {journal} {\bibinfo  {journal} {J. Electroanal.
  Chem.}\ }\textbf {\bibinfo {volume} {660}},\ \bibinfo {pages} {230} (\bibinfo
  {year} {2011})}\BibitemShut {NoStop}%
\bibitem [{\citenamefont {Nikitina}\ \emph
  {et~al.}(2014{\natexlab{a}})\citenamefont {Nikitina}, \citenamefont {Rudnev},
  \citenamefont {Tsirlina},\ and\ \citenamefont
  {Wandlowski}}]{Nikitina:2014ec}%
  \BibitemOpen
  \bibfield  {author} {\bibinfo {author} {\bibfnamefont {V.~A.}\ \bibnamefont
  {Nikitina}}, \bibinfo {author} {\bibfnamefont {A.~V.}\ \bibnamefont
  {Rudnev}}, \bibinfo {author} {\bibfnamefont {G.~A.}\ \bibnamefont
  {Tsirlina}}, \ and\ \bibinfo {author} {\bibfnamefont {T.}~\bibnamefont
  {Wandlowski}},\ }\href@noop {} {\bibfield  {journal} {\bibinfo  {journal} {J.
  Phys. Chem. C}\ }\textbf {\bibinfo {volume} {118}},\ \bibinfo {pages} {15970}
  (\bibinfo {year} {2014}{\natexlab{a}})}\BibitemShut {NoStop}%
\bibitem [{\citenamefont {Arzhantsev}\ \emph {et~al.}(2007)\citenamefont
  {Arzhantsev}, \citenamefont {Jin}, \citenamefont {Baker},\ and\ \citenamefont
  {Maroncelli}}]{Arzhantsev:2007ds}%
  \BibitemOpen
  \bibfield  {author} {\bibinfo {author} {\bibfnamefont {S.}~\bibnamefont
  {Arzhantsev}}, \bibinfo {author} {\bibfnamefont {H.}~\bibnamefont {Jin}},
  \bibinfo {author} {\bibfnamefont {G.~A.}\ \bibnamefont {Baker}}, \ and\
  \bibinfo {author} {\bibfnamefont {M.}~\bibnamefont {Maroncelli}},\
  }\href@noop {} {\bibfield  {journal} {\bibinfo  {journal} {Journal of
  Physical Chemistry B}\ }\textbf {\bibinfo {volume} {111}},\ \bibinfo {pages}
  {4978} (\bibinfo {year} {2007})}\BibitemShut {NoStop}%
\bibitem [{\citenamefont {Stoppa}\ \emph {et~al.}(2008)\citenamefont {Stoppa},
  \citenamefont {Hunger}, \citenamefont {Buchner}, \citenamefont {Hefter},
  \citenamefont {Thoman},\ and\ \citenamefont {Helm}}]{Stoppa:2008be}%
  \BibitemOpen
  \bibfield  {author} {\bibinfo {author} {\bibfnamefont {A.}~\bibnamefont
  {Stoppa}}, \bibinfo {author} {\bibfnamefont {J.}~\bibnamefont {Hunger}},
  \bibinfo {author} {\bibfnamefont {R.}~\bibnamefont {Buchner}}, \bibinfo
  {author} {\bibfnamefont {G.}~\bibnamefont {Hefter}}, \bibinfo {author}
  {\bibfnamefont {A.}~\bibnamefont {Thoman}}, \ and\ \bibinfo {author}
  {\bibfnamefont {H.}~\bibnamefont {Helm}},\ }\href@noop {} {\bibfield
  {journal} {\bibinfo  {journal} {J. Phys. Chem. B}\ }\textbf {\bibinfo
  {volume} {112}},\ \bibinfo {pages} {4854} (\bibinfo {year}
  {2008})}\BibitemShut {NoStop}%
\bibitem [{\citenamefont {Hunger}\ \emph {et~al.}(2009)\citenamefont {Hunger},
  \citenamefont {Stoppa}, \citenamefont {Schr~dle}, \citenamefont {Hefter},\
  and\ \citenamefont {Buchner}}]{Hunger:2009er}%
  \BibitemOpen
  \bibfield  {author} {\bibinfo {author} {\bibfnamefont {J.}~\bibnamefont
  {Hunger}}, \bibinfo {author} {\bibfnamefont {A.}~\bibnamefont {Stoppa}},
  \bibinfo {author} {\bibfnamefont {S.}~\bibnamefont {Schr~dle}}, \bibinfo
  {author} {\bibfnamefont {G.}~\bibnamefont {Hefter}}, \ and\ \bibinfo {author}
  {\bibfnamefont {R.}~\bibnamefont {Buchner}},\ }\href@noop {} {\bibfield
  {journal} {\bibinfo  {journal} {ChemPhysChem}\ }\textbf {\bibinfo {volume}
  {10}},\ \bibinfo {pages} {723} (\bibinfo {year} {2009})}\BibitemShut
  {NoStop}%
\bibitem [{\citenamefont {Weing{\"a}rtner}(2014)}]{Weingrtner:2014ib}%
  \BibitemOpen
  \bibfield  {author} {\bibinfo {author} {\bibfnamefont {H.}~\bibnamefont
  {Weing{\"a}rtner}},\ }\href@noop {} {\bibfield  {journal} {\bibinfo
  {journal} {J.\ Mol.\ Liq.}\ }\textbf {\bibinfo {volume} {192}},\ \bibinfo
  {pages} {185} (\bibinfo {year} {2014})}\BibitemShut {NoStop}%
\bibitem [{\citenamefont {Hong}\ \emph {et~al.}(2011)\citenamefont {Hong},
  \citenamefont {Smolin}, \citenamefont {Lindner}, \citenamefont {Sokolov},\
  and\ \citenamefont {Smith}}]{Hong:2011qf}%
  \BibitemOpen
  \bibfield  {author} {\bibinfo {author} {\bibfnamefont {L.}~\bibnamefont
  {Hong}}, \bibinfo {author} {\bibfnamefont {N.}~\bibnamefont {Smolin}},
  \bibinfo {author} {\bibfnamefont {B.}~\bibnamefont {Lindner}}, \bibinfo
  {author} {\bibfnamefont {A.~P.}\ \bibnamefont {Sokolov}}, \ and\ \bibinfo
  {author} {\bibfnamefont {J.~C.}\ \bibnamefont {Smith}},\ }\href@noop {}
  {\bibfield  {journal} {\bibinfo  {journal} {Phys. Rev. Lett.}\ }\textbf
  {\bibinfo {volume} {107}},\ \bibinfo {pages} {148102} (\bibinfo {year}
  {2011})}\BibitemShut {NoStop}%
\bibitem [{\citenamefont {Khodadadi}\ and\ \citenamefont
  {Sokolov}(2015)}]{Khodadadi:2015jp}%
  \BibitemOpen
  \bibfield  {author} {\bibinfo {author} {\bibfnamefont {S.}~\bibnamefont
  {Khodadadi}}\ and\ \bibinfo {author} {\bibfnamefont {A.~P.}\ \bibnamefont
  {Sokolov}},\ }\href@noop {} {\bibfield  {journal} {\bibinfo  {journal} {Soft
  Matter}\ }\textbf {\bibinfo {volume} {11}},\ \bibinfo {pages} {4984}
  (\bibinfo {year} {2015})}\BibitemShut {NoStop}%
\bibitem [{\citenamefont {Mondal}, \citenamefont {Mukherjee},\ and\
  \citenamefont {Bagchi}(2017)}]{Mondal:2017gf}%
  \BibitemOpen
  \bibfield  {author} {\bibinfo {author} {\bibfnamefont {S.}~\bibnamefont
  {Mondal}}, \bibinfo {author} {\bibfnamefont {S.}~\bibnamefont {Mukherjee}}, \
  and\ \bibinfo {author} {\bibfnamefont {B.}~\bibnamefont {Bagchi}},\
  }\href@noop {} {\bibfield  {journal} {\bibinfo  {journal} {Chem. Phys.
  Lett.}\ ,\ \bibinfo {pages} {1}} (\bibinfo {year} {2017})}\BibitemShut
  {NoStop}%
\bibitem [{\citenamefont {Martin}\ and\ \citenamefont
  {Matyushov}(2015)}]{DMjcp1:15}%
  \BibitemOpen
  \bibfield  {author} {\bibinfo {author} {\bibfnamefont {D.~R.}\ \bibnamefont
  {Martin}}\ and\ \bibinfo {author} {\bibfnamefont {D.~V.}\ \bibnamefont
  {Matyushov}},\ }\href@noop {} {\bibfield  {journal} {\bibinfo  {journal} {J.
  Chem. Phys.}\ }\textbf {\bibinfo {volume} {142}},\ \bibinfo {pages} {161101}
  (\bibinfo {year} {2015})}\BibitemShut {NoStop}%
\bibitem [{\citenamefont {Ghorai}\ and\ \citenamefont
  {Matyushov}(2006)}]{DMjpcb1:06}%
  \BibitemOpen
  \bibfield  {author} {\bibinfo {author} {\bibfnamefont {P.~K.}\ \bibnamefont
  {Ghorai}}\ and\ \bibinfo {author} {\bibfnamefont {D.~V.}\ \bibnamefont
  {Matyushov}},\ }\href@noop {} {\bibfield  {journal} {\bibinfo  {journal} {J.
  Phys. Chem. B}\ }\textbf {\bibinfo {volume} {110}},\ \bibinfo {pages} {1866}
  (\bibinfo {year} {2006})}\BibitemShut {NoStop}%
\bibitem [{\citenamefont {Richert}(2002)}]{Richert:02}%
  \BibitemOpen
  \bibfield  {author} {\bibinfo {author} {\bibfnamefont {R.}~\bibnamefont
  {Richert}},\ }\href@noop {} {\bibfield  {journal} {\bibinfo  {journal} {J.
  Phys.: Condens. Matter}\ }\textbf {\bibinfo {volume} {14}},\ \bibinfo {pages}
  {R703} (\bibinfo {year} {2002})}\BibitemShut {NoStop}%
\bibitem [{\citenamefont {Richert}(2015)}]{Richert:2014wa}%
  \BibitemOpen
  \bibfield  {author} {\bibinfo {author} {\bibfnamefont {R.}~\bibnamefont
  {Richert}},\ }\href@noop {} {\bibfield  {journal} {\bibinfo  {journal} {Adv.
  Chem. Phys.}\ }\textbf {\bibinfo {volume} {156}},\ \bibinfo {pages} {101}
  (\bibinfo {year} {2015})}\BibitemShut {NoStop}%
\bibitem [{\citenamefont {Bagchi}\ and\ \citenamefont
  {Chandra}(1991)}]{Bagchi:91}%
  \BibitemOpen
  \bibfield  {author} {\bibinfo {author} {\bibfnamefont {B.}~\bibnamefont
  {Bagchi}}\ and\ \bibinfo {author} {\bibfnamefont {A.}~\bibnamefont
  {Chandra}},\ }\href@noop {} {\bibfield  {journal} {\bibinfo  {journal} {Adv.
  Chem. Phys.}\ }\textbf {\bibinfo {volume} {80}},\ \bibinfo {pages} {1}
  (\bibinfo {year} {1991})}\BibitemShut {NoStop}%
\bibitem [{\citenamefont {Chen}\ and\ \citenamefont {Meyer}(1996)}]{Chen:96}%
  \BibitemOpen
  \bibfield  {author} {\bibinfo {author} {\bibfnamefont {P.}~\bibnamefont
  {Chen}}\ and\ \bibinfo {author} {\bibfnamefont {T.~J.}\ \bibnamefont
  {Meyer}},\ }\href@noop {} {\bibfield  {journal} {\bibinfo  {journal} {Inorg.
  Chem.}\ }\textbf {\bibinfo {volume} {35}},\ \bibinfo {pages} {5520} (\bibinfo
  {year} {1996})}\BibitemShut {NoStop}%
\bibitem [{\citenamefont {G{\"o}rlach}\ \emph {et~al.}(1995)\citenamefont
  {G{\"o}rlach}, \citenamefont {Gydax}, \citenamefont {Lubini},\ and\
  \citenamefont {Wild}}]{Gorlach:95}%
  \BibitemOpen
  \bibfield  {author} {\bibinfo {author} {\bibfnamefont {E.}~\bibnamefont
  {G{\"o}rlach}}, \bibinfo {author} {\bibfnamefont {H.}~\bibnamefont {Gydax}},
  \bibinfo {author} {\bibfnamefont {P.}~\bibnamefont {Lubini}}, \ and\ \bibinfo
  {author} {\bibfnamefont {U.~P.}\ \bibnamefont {Wild}},\ }\href@noop {}
  {\bibfield  {journal} {\bibinfo  {journal} {Chem. Phys.}\ }\textbf {\bibinfo
  {volume} {194}},\ \bibinfo {pages} {185} (\bibinfo {year}
  {1995})}\BibitemShut {NoStop}%
\bibitem [{\citenamefont {Richert}(2000)}]{Richert:2000wq}%
  \BibitemOpen
  \bibfield  {author} {\bibinfo {author} {\bibfnamefont {R.}~\bibnamefont
  {Richert}},\ }\href@noop {} {\bibfield  {journal} {\bibinfo  {journal} {J.
  Chem. Phys.}\ }\textbf {\bibinfo {volume} {113}},\ \bibinfo {pages} {8404}
  (\bibinfo {year} {2000})}\BibitemShut {NoStop}%
\bibitem [{\citenamefont {Goes}\ \emph {et~al.}(2002)\citenamefont {Goes},
  \citenamefont {de~Groot}, \citenamefont {Koeberg}, \citenamefont {Verhoeven},
  \citenamefont {Lokan}, \citenamefont {Shephard},\ and\ \citenamefont
  {Paddon-Row}}]{Goes:02}%
  \BibitemOpen
  \bibfield  {author} {\bibinfo {author} {\bibfnamefont {M.}~\bibnamefont
  {Goes}}, \bibinfo {author} {\bibfnamefont {M.}~\bibnamefont {de~Groot}},
  \bibinfo {author} {\bibfnamefont {M.}~\bibnamefont {Koeberg}}, \bibinfo
  {author} {\bibfnamefont {J.~W.}\ \bibnamefont {Verhoeven}}, \bibinfo {author}
  {\bibfnamefont {N.~R.}\ \bibnamefont {Lokan}}, \bibinfo {author}
  {\bibfnamefont {M.~J.}\ \bibnamefont {Shephard}}, \ and\ \bibinfo {author}
  {\bibfnamefont {M.~N.}\ \bibnamefont {Paddon-Row}},\ }\href@noop {}
  {\bibfield  {journal} {\bibinfo  {journal} {J. Phys. Chem. A}\ }\textbf
  {\bibinfo {volume} {106}},\ \bibinfo {pages} {2129} (\bibinfo {year}
  {2002})}\BibitemShut {NoStop}%
\bibitem [{\citenamefont {Lakowicz}(2000)}]{Lakowicz:2000jn}%
  \BibitemOpen
  \bibfield  {author} {\bibinfo {author} {\bibfnamefont {J.~R.}\ \bibnamefont
  {Lakowicz}},\ }\href@noop {} {\bibfield  {journal} {\bibinfo  {journal}
  {Photochemistry and Photobiology}\ }\textbf {\bibinfo {volume} {72}},\
  \bibinfo {pages} {421} (\bibinfo {year} {2000})}\BibitemShut {NoStop}%
\bibitem [{\citenamefont {Matyushov}(2013)}]{DMjcp2:13}%
  \BibitemOpen
  \bibfield  {author} {\bibinfo {author} {\bibfnamefont {D.~V.}\ \bibnamefont
  {Matyushov}},\ }\href@noop {} {\bibfield  {journal} {\bibinfo  {journal} {J.
  Chem. Phys.}\ }\textbf {\bibinfo {volume} {139}},\ \bibinfo {pages} {025102}
  (\bibinfo {year} {2013})}\BibitemShut {NoStop}%
\bibitem [{\citenamefont {Qin}\ \emph {et~al.}(2017)\citenamefont {Qin},
  \citenamefont {Zhang}, \citenamefont {Wang},\ and\ \citenamefont
  {Zhong}}]{Qin:2017kk}%
  \BibitemOpen
  \bibfield  {author} {\bibinfo {author} {\bibfnamefont {Y.}~\bibnamefont
  {Qin}}, \bibinfo {author} {\bibfnamefont {L.}~\bibnamefont {Zhang}}, \bibinfo
  {author} {\bibfnamefont {L.}~\bibnamefont {Wang}}, \ and\ \bibinfo {author}
  {\bibfnamefont {D.}~\bibnamefont {Zhong}},\ }\href@noop {} {\bibfield
  {journal} {\bibinfo  {journal} {J. Phys. Chem. Lett.}\ }\textbf {\bibinfo
  {volume} {8}},\ \bibinfo {pages} {1124} (\bibinfo {year} {2017})}\BibitemShut
  {NoStop}%
\bibitem [{\citenamefont {Jimenez}\ \emph {et~al.}(1994)\citenamefont
  {Jimenez}, \citenamefont {Fleming}, \citenamefont {Kumar},\ and\
  \citenamefont {Maroncelli}}]{Jimenez:94}%
  \BibitemOpen
  \bibfield  {author} {\bibinfo {author} {\bibfnamefont {R.}~\bibnamefont
  {Jimenez}}, \bibinfo {author} {\bibfnamefont {G.~R.}\ \bibnamefont
  {Fleming}}, \bibinfo {author} {\bibfnamefont {P.~V.}\ \bibnamefont {Kumar}},
  \ and\ \bibinfo {author} {\bibfnamefont {M.}~\bibnamefont {Maroncelli}},\
  }\href@noop {} {\bibfield  {journal} {\bibinfo  {journal} {Nature}\ }\textbf
  {\bibinfo {volume} {369}},\ \bibinfo {pages} {471} (\bibinfo {year}
  {1994})}\BibitemShut {NoStop}%
\bibitem [{\citenamefont {Bagchi}\ and\ \citenamefont
  {Gayathri}(1999)}]{BagchiGayathri:99}%
  \BibitemOpen
  \bibfield  {author} {\bibinfo {author} {\bibfnamefont {B.}~\bibnamefont
  {Bagchi}}\ and\ \bibinfo {author} {\bibfnamefont {N.}~\bibnamefont
  {Gayathri}},\ }\href@noop {} {\bibfield  {journal} {\bibinfo  {journal} {Adv.
  Chem. Phys.}\ }\textbf {\bibinfo {volume} {107}},\ \bibinfo {pages} {1}
  (\bibinfo {year} {1999})}\BibitemShut {NoStop}%
\bibitem [{\citenamefont {Ediger}(2000)}]{Ediger:00}%
  \BibitemOpen
  \bibfield  {author} {\bibinfo {author} {\bibfnamefont {M.~D.}\ \bibnamefont
  {Ediger}},\ }\href@noop {} {\bibfield  {journal} {\bibinfo  {journal} {Annu.
  Rev. Phys. Chem.}\ }\textbf {\bibinfo {volume} {51}},\ \bibinfo {pages} {99}
  (\bibinfo {year} {2000})}\BibitemShut {NoStop}%
\bibitem [{\citenamefont {van~der Zwan}\ and\ \citenamefont
  {Hynes}(1985)}]{Zwan:85}%
  \BibitemOpen
  \bibfield  {author} {\bibinfo {author} {\bibfnamefont {G.}~\bibnamefont
  {van~der Zwan}}\ and\ \bibinfo {author} {\bibfnamefont {J.~T.}\ \bibnamefont
  {Hynes}},\ }\href@noop {} {\bibfield  {journal} {\bibinfo  {journal} {J.\
  Phys.\ Chem.}\ }\textbf {\bibinfo {volume} {89}},\ \bibinfo {pages} {4181}
  (\bibinfo {year} {1985})}\BibitemShut {NoStop}%
\bibitem [{\citenamefont {Carter}\ and\ \citenamefont
  {Hynes}(1991)}]{Carter:91}%
  \BibitemOpen
  \bibfield  {author} {\bibinfo {author} {\bibfnamefont {E.~A.}\ \bibnamefont
  {Carter}}\ and\ \bibinfo {author} {\bibfnamefont {J.~T.}\ \bibnamefont
  {Hynes}},\ }\href@noop {} {\bibfield  {journal} {\bibinfo  {journal} {J.\
  Chem.\ Phys.}\ }\textbf {\bibinfo {volume} {94}},\ \bibinfo {pages} {5961}
  (\bibinfo {year} {1991})}\BibitemShut {NoStop}%
\bibitem [{\citenamefont {Reynolds}\ \emph {et~al.}(1996)\citenamefont
  {Reynolds}, \citenamefont {Gardecki}, \citenamefont {Frankland},\ and\
  \citenamefont {Maroncelli}}]{Reynolds:96}%
  \BibitemOpen
  \bibfield  {author} {\bibinfo {author} {\bibfnamefont {L.}~\bibnamefont
  {Reynolds}}, \bibinfo {author} {\bibfnamefont {J.~A.}\ \bibnamefont
  {Gardecki}}, \bibinfo {author} {\bibfnamefont {S.~J.~V.}\ \bibnamefont
  {Frankland}}, \ and\ \bibinfo {author} {\bibfnamefont {M.}~\bibnamefont
  {Maroncelli}},\ }\href@noop {} {\bibfield  {journal} {\bibinfo  {journal} {J.
  Phys. Chem.}\ }\textbf {\bibinfo {volume} {100}},\ \bibinfo {pages} {10337}
  (\bibinfo {year} {1996})}\BibitemShut {NoStop}%
\bibitem [{\citenamefont {Mukamel}(1995)}]{Mukamel:95}%
  \BibitemOpen
  \bibfield  {author} {\bibinfo {author} {\bibfnamefont {S.}~\bibnamefont
  {Mukamel}},\ }\href@noop {} {\emph {\bibinfo {title} {Principles of Nonlinear
  Optical Spectroscopy}}}\ (\bibinfo  {publisher} {Oxford University Press},\
  \bibinfo {address} {New York},\ \bibinfo {year} {1995})\BibitemShut {NoStop}%
\bibitem [{\citenamefont {Matyushov}(2005)}]{DMjcp2:05}%
  \BibitemOpen
  \bibfield  {author} {\bibinfo {author} {\bibfnamefont {D.~V.}\ \bibnamefont
  {Matyushov}},\ }\href@noop {} {\bibfield  {journal} {\bibinfo  {journal} {J.\
  Chem.\ Phys.}\ }\textbf {\bibinfo {volume} {122}},\ \bibinfo {pages} {084507}
  (\bibinfo {year} {2005})}\BibitemShut {NoStop}%
\bibitem [{\citenamefont {Matyushov}(2015)}]{DMjpcm:15}%
  \BibitemOpen
  \bibfield  {author} {\bibinfo {author} {\bibfnamefont {D.~V.}\ \bibnamefont
  {Matyushov}},\ }\href@noop {} {\bibfield  {journal} {\bibinfo  {journal} {J.
  Phys.: Condens. Matter}\ }\textbf {\bibinfo {volume} {27}},\ \bibinfo {pages}
  {473001} (\bibinfo {year} {2015})}\BibitemShut {NoStop}%
\bibitem [{\citenamefont {Marcus}\ and\ \citenamefont
  {Sutin}(1985)}]{MarcusSutin}%
  \BibitemOpen
  \bibfield  {author} {\bibinfo {author} {\bibfnamefont {R.~A.}\ \bibnamefont
  {Marcus}}\ and\ \bibinfo {author} {\bibfnamefont {N.}~\bibnamefont {Sutin}},\
  }\href@noop {} {\bibfield  {journal} {\bibinfo  {journal} {Biochim. Biophys.
  Acta}\ }\textbf {\bibinfo {volume} {811}},\ \bibinfo {pages} {265} (\bibinfo
  {year} {1985})}\BibitemShut {NoStop}%
\bibitem [{\citenamefont {Hale}(1968)}]{Hale:1968ju}%
  \BibitemOpen
  \bibfield  {author} {\bibinfo {author} {\bibfnamefont {J.~M.}\ \bibnamefont
  {Hale}},\ }\href@noop {} {\bibfield  {journal} {\bibinfo  {journal} {J.
  Electroanal. Chem.}\ }\textbf {\bibinfo {volume} {19}},\ \bibinfo {pages}
  {315} (\bibinfo {year} {1968})}\BibitemShut {NoStop}%
\bibitem [{\citenamefont {Oldham}\ and\ \citenamefont
  {Myland}(2011)}]{Oldham:2011fk}%
  \BibitemOpen
  \bibfield  {author} {\bibinfo {author} {\bibfnamefont {K.~B.}\ \bibnamefont
  {Oldham}}\ and\ \bibinfo {author} {\bibfnamefont {J.~C.}\ \bibnamefont
  {Myland}},\ }\href {\doibase
  http://dx.doi.org/10.1016/j.jelechem.2011.01.044} {\bibfield  {journal}
  {\bibinfo  {journal} {Journal of Electroanalytical Chemistry}\ }\textbf
  {\bibinfo {volume} {655}},\ \bibinfo {pages} {65} (\bibinfo {year}
  {2011})}\BibitemShut {NoStop}%
\bibitem [{\citenamefont {Newton}\ and\ \citenamefont
  {Smalley}(2007)}]{Newton:2007fk}%
  \BibitemOpen
  \bibfield  {author} {\bibinfo {author} {\bibfnamefont {M.~D.}\ \bibnamefont
  {Newton}}\ and\ \bibinfo {author} {\bibfnamefont {J.~F.}\ \bibnamefont
  {Smalley}},\ }\href {\doibase 10.1039/B611448B} {\bibfield  {journal}
  {\bibinfo  {journal} {Phys. Chem. Chem. Phys.}\ }\textbf {\bibinfo {volume}
  {9}},\ \bibinfo {pages} {555} (\bibinfo {year} {2007})}\BibitemShut {NoStop}%
\bibitem [{\citenamefont {Abramowitz}\ and\ \citenamefont
  {Stegun}(1972)}]{Abramowitz:72}%
  \BibitemOpen
  \bibinfo {editor} {\bibfnamefont {M.}~\bibnamefont {Abramowitz}}\ and\
  \bibinfo {editor} {\bibfnamefont {I.~A.}\ \bibnamefont {Stegun}},\ eds.,\
  \href@noop {} {\emph {\bibinfo {title} {Handbook of Mathematical
  Functions}}}\ (\bibinfo  {publisher} {Dover},\ \bibinfo {address} {New
  York},\ \bibinfo {year} {1972})\BibitemShut {NoStop}%
\bibitem [{\citenamefont {Compton}\ and\ \citenamefont
  {Banks}(2009)}]{ComptonBanks}%
  \BibitemOpen
  \bibfield  {author} {\bibinfo {author} {\bibfnamefont {R.~G.}\ \bibnamefont
  {Compton}}\ and\ \bibinfo {author} {\bibfnamefont {C.~E.}\ \bibnamefont
  {Banks}},\ }\href@noop {} {\emph {\bibinfo {title} {Understanding
  {V}oltammetry}}}\ (\bibinfo  {publisher} {World Scientific},\ \bibinfo
  {address} {Singapore},\ \bibinfo {year} {2009})\BibitemShut {NoStop}%
\bibitem [{\citenamefont {Yue}\ \emph {et~al.}(2006)\citenamefont {Yue},
  \citenamefont {Khoshtariya}, \citenamefont {Waldeck}, \citenamefont
  {Grochol}, \citenamefont {Hildebrandt},\ and\ \citenamefont
  {Murgida}}]{Yue:2006wo}%
  \BibitemOpen
  \bibfield  {author} {\bibinfo {author} {\bibfnamefont {H.}~\bibnamefont
  {Yue}}, \bibinfo {author} {\bibfnamefont {D.}~\bibnamefont {Khoshtariya}},
  \bibinfo {author} {\bibfnamefont {D.~H.}\ \bibnamefont {Waldeck}}, \bibinfo
  {author} {\bibfnamefont {J.}~\bibnamefont {Grochol}}, \bibinfo {author}
  {\bibfnamefont {P.}~\bibnamefont {Hildebrandt}}, \ and\ \bibinfo {author}
  {\bibfnamefont {D.~H.}\ \bibnamefont {Murgida}},\ }\bibfield  {booktitle}
  {\emph {\bibinfo {booktitle} {The Journal of Physical Chemistry B}},\ }\href
  {\doibase 10.1021/jp0620670} {\bibfield  {journal} {\bibinfo  {journal} {J.
  Phys. Chem. B}\ }\textbf {\bibinfo {volume} {110}},\ \bibinfo {pages} {19906}
  (\bibinfo {year} {2006})}\BibitemShut {NoStop}%
\bibitem [{\citenamefont {Becka}\ and\ \citenamefont
  {Miller}(1992)}]{Becka:1992mz}%
  \BibitemOpen
  \bibfield  {author} {\bibinfo {author} {\bibfnamefont {A.~M.}\ \bibnamefont
  {Becka}}\ and\ \bibinfo {author} {\bibfnamefont {C.~J.}\ \bibnamefont
  {Miller}},\ }\href {\doibase 10.1021/j100185a049} {\bibfield  {journal}
  {\bibinfo  {journal} {J. Phys. Chem.}\ }\textbf {\bibinfo {volume} {96}},\
  \bibinfo {pages} {2657} (\bibinfo {year} {1992})}\BibitemShut {NoStop}%
\bibitem [{\citenamefont {Terrettaz}, \citenamefont {Cheng},\ and\
  \citenamefont {Miller}(1996)}]{Terrettaz:1996gv}%
  \BibitemOpen
  \bibfield  {author} {\bibinfo {author} {\bibfnamefont {S.}~\bibnamefont
  {Terrettaz}}, \bibinfo {author} {\bibfnamefont {J.}~\bibnamefont {Cheng}}, \
  and\ \bibinfo {author} {\bibfnamefont {C.~J.}\ \bibnamefont {Miller}},\
  }\href@noop {} {\bibfield  {journal} {\bibinfo  {journal} {J. Am. Chem.
  Soc.}\ }\textbf {\bibinfo {volume} {118}},\ \bibinfo {pages} {7857} (\bibinfo
  {year} {1996})}\BibitemShut {NoStop}%
\bibitem [{\citenamefont {Miller}(1995)}]{Miller:95}%
  \BibitemOpen
  \bibfield  {author} {\bibinfo {author} {\bibfnamefont {C.~J.}\ \bibnamefont
  {Miller}},\ }\enquote {\bibinfo {title} {Physical electrochemistry:
  {P}rinciples, methods, and applications},}\ \ (\bibinfo  {publisher} {Marcell
  Dekker},\ \bibinfo {year} {1995})\ Chap.\ \bibinfo {chapter} {Heterogeneous
  electron transfer kinetics at metallic electrodes}, pp.\ \bibinfo {pages}
  {27--79}\BibitemShut {NoStop}%
\bibitem [{\citenamefont {Laviron}(1979)}]{Laviron79}%
  \BibitemOpen
  \bibfield  {author} {\bibinfo {author} {\bibfnamefont {E.}~\bibnamefont
  {Laviron}},\ }\href@noop {} {\bibfield  {journal} {\bibinfo  {journal} {J.
  Electroanal. Chem.}\ }\textbf {\bibinfo {volume} {101}},\ \bibinfo {pages}
  {19} (\bibinfo {year} {1979})}\BibitemShut {NoStop}%
\bibitem [{\citenamefont {Honeychurch}(1999)}]{Honeychurch:1999uq}%
  \BibitemOpen
  \bibfield  {author} {\bibinfo {author} {\bibfnamefont {M.~J.}\ \bibnamefont
  {Honeychurch}},\ }\bibfield  {booktitle} {\emph {\bibinfo {booktitle}
  {Langmuir}},\ }\href {\doibase 10.1021/la990169u} {\bibfield  {journal}
  {\bibinfo  {journal} {Langmuir}\ }\textbf {\bibinfo {volume} {15}},\ \bibinfo
  {pages} {5158} (\bibinfo {year} {1999})}\BibitemShut {NoStop}%
\bibitem [{\citenamefont {Landau}\ and\ \citenamefont
  {Lifshitz}(1984)}]{Landau8}%
  \BibitemOpen
  \bibfield  {author} {\bibinfo {author} {\bibfnamefont {L.~D.}\ \bibnamefont
  {Landau}}\ and\ \bibinfo {author} {\bibfnamefont {E.~M.}\ \bibnamefont
  {Lifshitz}},\ }\href@noop {} {\emph {\bibinfo {title} {Electrodynamics of
  {C}ontinuous {M}edia}}}\ (\bibinfo  {publisher} {Pergamon},\ \bibinfo
  {address} {Oxford},\ \bibinfo {year} {1984})\BibitemShut {NoStop}%
\bibitem [{\citenamefont
  {Lynden-Bell}(2007{\natexlab{a}})}]{Lynden-Bell:2007pd}%
  \BibitemOpen
  \bibfield  {author} {\bibinfo {author} {\bibfnamefont {R.~M.}\ \bibnamefont
  {Lynden-Bell}},\ }\bibfield  {booktitle} {\emph {\bibinfo {booktitle} {The
  Journal of Physical Chemistry B}},\ }\href {\doibase 10.1021/jp074298s}
  {\bibfield  {journal} {\bibinfo  {journal} {J. Phys. Chem. B}\ }\textbf
  {\bibinfo {volume} {111}},\ \bibinfo {pages} {10800} (\bibinfo {year}
  {2007}{\natexlab{a}})}\BibitemShut {NoStop}%
\bibitem [{\citenamefont
  {Lynden-Bell}(2007{\natexlab{b}})}]{LyndenBell:2007ff}%
  \BibitemOpen
  \bibfield  {author} {\bibinfo {author} {\bibfnamefont {R.~M.}\ \bibnamefont
  {Lynden-Bell}},\ }\href@noop {} {\bibfield  {journal} {\bibinfo  {journal}
  {Electrochemistry Communications}\ }\textbf {\bibinfo {volume} {9}},\
  \bibinfo {pages} {1857} (\bibinfo {year} {2007}{\natexlab{b}})}\BibitemShut
  {NoStop}%
\bibitem [{\citenamefont {Sonnleitner}\ \emph {et~al.}(2015)\citenamefont
  {Sonnleitner}, \citenamefont {Turton}, \citenamefont {Hefter}, \citenamefont
  {Ortner}, \citenamefont {Waselikowski}, \citenamefont {Walther},
  \citenamefont {Wynne},\ and\ \citenamefont {Buchner}}]{Sonnleitner:2015fla}%
  \BibitemOpen
  \bibfield  {author} {\bibinfo {author} {\bibfnamefont {T.}~\bibnamefont
  {Sonnleitner}}, \bibinfo {author} {\bibfnamefont {D.~A.}\ \bibnamefont
  {Turton}}, \bibinfo {author} {\bibfnamefont {G.}~\bibnamefont {Hefter}},
  \bibinfo {author} {\bibfnamefont {A.}~\bibnamefont {Ortner}}, \bibinfo
  {author} {\bibfnamefont {S.}~\bibnamefont {Waselikowski}}, \bibinfo {author}
  {\bibfnamefont {M.}~\bibnamefont {Walther}}, \bibinfo {author} {\bibfnamefont
  {K.}~\bibnamefont {Wynne}}, \ and\ \bibinfo {author} {\bibfnamefont
  {R.}~\bibnamefont {Buchner}},\ }\href@noop {} {\bibfield  {journal} {\bibinfo
   {journal} {J. Phys. Chem. B}\ }\textbf {\bibinfo {volume} {119}},\ \bibinfo
  {pages} {8826} (\bibinfo {year} {2015})}\BibitemShut {NoStop}%
\bibitem [{\citenamefont {Leys}\ \emph {et~al.}(2008)\citenamefont {Leys},
  \citenamefont {W~bbenhorst}, \citenamefont {Preethy~Menon}, \citenamefont
  {Rajesh}, \citenamefont {Thoen}, \citenamefont {Glorieux}, \citenamefont
  {Nockemann}, \citenamefont {Thijs}, \citenamefont {Binnemans},\ and\
  \citenamefont {Longuemart}}]{Leys:2008il}%
  \BibitemOpen
  \bibfield  {author} {\bibinfo {author} {\bibfnamefont {J.}~\bibnamefont
  {Leys}}, \bibinfo {author} {\bibfnamefont {M.}~\bibnamefont {W~bbenhorst}},
  \bibinfo {author} {\bibfnamefont {C.}~\bibnamefont {Preethy~Menon}}, \bibinfo
  {author} {\bibfnamefont {R.}~\bibnamefont {Rajesh}}, \bibinfo {author}
  {\bibfnamefont {J.}~\bibnamefont {Thoen}}, \bibinfo {author} {\bibfnamefont
  {C.}~\bibnamefont {Glorieux}}, \bibinfo {author} {\bibfnamefont
  {P.}~\bibnamefont {Nockemann}}, \bibinfo {author} {\bibfnamefont
  {B.}~\bibnamefont {Thijs}}, \bibinfo {author} {\bibfnamefont
  {K.}~\bibnamefont {Binnemans}}, \ and\ \bibinfo {author} {\bibfnamefont
  {S.~p.}\ \bibnamefont {Longuemart}},\ }\href@noop {} {\bibfield  {journal}
  {\bibinfo  {journal} {J. Chem. Phys.}\ }\textbf {\bibinfo {volume} {128}},\
  \bibinfo {pages} {064509} (\bibinfo {year} {2008})}\BibitemShut {NoStop}%
\bibitem [{\citenamefont {Serghei}\ \emph {et~al.}(2009)\citenamefont
  {Serghei}, \citenamefont {Tress}, \citenamefont {Sangoro},\ and\
  \citenamefont {Kremer}}]{Serghei:2009dl}%
  \BibitemOpen
  \bibfield  {author} {\bibinfo {author} {\bibfnamefont {A.}~\bibnamefont
  {Serghei}}, \bibinfo {author} {\bibfnamefont {M.}~\bibnamefont {Tress}},
  \bibinfo {author} {\bibfnamefont {J.~R.}\ \bibnamefont {Sangoro}}, \ and\
  \bibinfo {author} {\bibfnamefont {F.}~\bibnamefont {Kremer}},\ }\href@noop {}
  {\bibfield  {journal} {\bibinfo  {journal} {Phys. Rev. B}\ }\textbf {\bibinfo
  {volume} {80}},\ \bibinfo {pages} {654} (\bibinfo {year} {2009})}\BibitemShut
  {NoStop}%
\bibitem [{\citenamefont {Ito}\ and\ \citenamefont
  {Richert}(2007)}]{Ito:2007tw}%
  \BibitemOpen
  \bibfield  {author} {\bibinfo {author} {\bibfnamefont {N.}~\bibnamefont
  {Ito}}\ and\ \bibinfo {author} {\bibfnamefont {R.}~\bibnamefont {Richert}},\
  }\href@noop {} {\bibfield  {journal} {\bibinfo  {journal} {J. Phys. Chem. B}\
  }\textbf {\bibinfo {volume} {111}},\ \bibinfo {pages} {5016} (\bibinfo {year}
  {2007})}\BibitemShut {NoStop}%
\bibitem [{\citenamefont {Wang}\ \emph {et~al.}(2007)\citenamefont {Wang},
  \citenamefont {Jiang}, \citenamefont {Yan},\ and\ \citenamefont
  {Voth}}]{Wang:2007iy}%
  \BibitemOpen
  \bibfield  {author} {\bibinfo {author} {\bibfnamefont {Y.}~\bibnamefont
  {Wang}}, \bibinfo {author} {\bibfnamefont {W.}~\bibnamefont {Jiang}},
  \bibinfo {author} {\bibfnamefont {T.}~\bibnamefont {Yan}}, \ and\ \bibinfo
  {author} {\bibfnamefont {G.~A.}\ \bibnamefont {Voth}},\ }\href@noop {}
  {\bibfield  {journal} {\bibinfo  {journal} {Acc. Chem. Res.}\ }\textbf
  {\bibinfo {volume} {40}},\ \bibinfo {pages} {1193} (\bibinfo {year}
  {2007})}\BibitemShut {NoStop}%
\bibitem [{\citenamefont {Fayer}(2014)}]{Fayer:2014et}%
  \BibitemOpen
  \bibfield  {author} {\bibinfo {author} {\bibfnamefont {M.~D.}\ \bibnamefont
  {Fayer}},\ }\href@noop {} {\bibfield  {journal} {\bibinfo  {journal} {Chem.
  Phys. Lett.}\ }\textbf {\bibinfo {volume} {616-617}},\ \bibinfo {pages} {259}
  (\bibinfo {year} {2014})}\BibitemShut {NoStop}%
\bibitem [{\citenamefont {Shim}\ and\ \citenamefont {Kim}(2009)}]{Shim:2009jf}%
  \BibitemOpen
  \bibfield  {author} {\bibinfo {author} {\bibfnamefont {Y.}~\bibnamefont
  {Shim}}\ and\ \bibinfo {author} {\bibfnamefont {H.~J.}\ \bibnamefont {Kim}},\
  }\href@noop {} {\bibfield  {journal} {\bibinfo  {journal} {J. Phys. Chem. B}\
  }\textbf {\bibinfo {volume} {113}},\ \bibinfo {pages} {12964} (\bibinfo
  {year} {2009})}\BibitemShut {NoStop}%
\bibitem [{\citenamefont {Mladenova}\ \emph {et~al.}(2016)\citenamefont
  {Mladenova}, \citenamefont {Kattnig}, \citenamefont {Sudy}, \citenamefont
  {Choto},\ and\ \citenamefont {Grampp}}]{Mladenova:2016gu}%
  \BibitemOpen
  \bibfield  {author} {\bibinfo {author} {\bibfnamefont {B.~Y.}\ \bibnamefont
  {Mladenova}}, \bibinfo {author} {\bibfnamefont {D.~R.}\ \bibnamefont
  {Kattnig}}, \bibinfo {author} {\bibfnamefont {B.}~\bibnamefont {Sudy}},
  \bibinfo {author} {\bibfnamefont {P.}~\bibnamefont {Choto}}, \ and\ \bibinfo
  {author} {\bibfnamefont {G.}~\bibnamefont {Grampp}},\ }\href@noop {}
  {\bibfield  {journal} {\bibinfo  {journal} {Phys. Chem. Chem. Phys.}\
  }\textbf {\bibinfo {volume} {18}},\ \bibinfo {pages} {14442} (\bibinfo {year}
  {2016})}\BibitemShut {NoStop}%
\bibitem [{\citenamefont {Kashyap}\ and\ \citenamefont
  {Biswas}(2008)}]{Kashyap:2008fe}%
  \BibitemOpen
  \bibfield  {author} {\bibinfo {author} {\bibfnamefont {H.~K.}\ \bibnamefont
  {Kashyap}}\ and\ \bibinfo {author} {\bibfnamefont {R.}~\bibnamefont
  {Biswas}},\ }\href@noop {} {\bibfield  {journal} {\bibinfo  {journal} {J.
  Phys. Chem. B}\ }\textbf {\bibinfo {volume} {112}},\ \bibinfo {pages} {12431}
  (\bibinfo {year} {2008})}\BibitemShut {NoStop}%
\bibitem [{\citenamefont {Roy}\ and\ \citenamefont
  {Maroncelli}(2012)}]{Roy:2012gp}%
  \BibitemOpen
  \bibfield  {author} {\bibinfo {author} {\bibfnamefont {D.}~\bibnamefont
  {Roy}}\ and\ \bibinfo {author} {\bibfnamefont {M.}~\bibnamefont
  {Maroncelli}},\ }\href@noop {} {\bibfield  {journal} {\bibinfo  {journal} {J.
  Phys. Chem. B}\ }\textbf {\bibinfo {volume} {116}},\ \bibinfo {pages} {5951}
  (\bibinfo {year} {2012})}\BibitemShut {NoStop}%
\bibitem [{\citenamefont {Terranova}\ and\ \citenamefont
  {Corcelli}(2013)}]{Terranova:2013dj}%
  \BibitemOpen
  \bibfield  {author} {\bibinfo {author} {\bibfnamefont {Z.~L.}\ \bibnamefont
  {Terranova}}\ and\ \bibinfo {author} {\bibfnamefont {S.~A.}\ \bibnamefont
  {Corcelli}},\ }\href@noop {} {\bibfield  {journal} {\bibinfo  {journal} {J.
  Phys. Chem. B}\ }\textbf {\bibinfo {volume} {117}},\ \bibinfo {pages} {15659}
  (\bibinfo {year} {2013})}\BibitemShut {NoStop}%
\bibitem [{\citenamefont {Matyushov}(1993)}]{DMmp:93}%
  \BibitemOpen
  \bibfield  {author} {\bibinfo {author} {\bibfnamefont {D.~V.}\ \bibnamefont
  {Matyushov}},\ }\href@noop {} {\bibfield  {journal} {\bibinfo  {journal}
  {Mol. Phys.}\ }\textbf {\bibinfo {volume} {79}},\ \bibinfo {pages} {795}
  (\bibinfo {year} {1993})}\BibitemShut {NoStop}%
\bibitem [{\citenamefont {Dinpajooh}, \citenamefont {Newton},\ and\
  \citenamefont {Matyushov}(2017)}]{DMjcp1:17}%
  \BibitemOpen
  \bibfield  {author} {\bibinfo {author} {\bibfnamefont {M.}~\bibnamefont
  {Dinpajooh}}, \bibinfo {author} {\bibfnamefont {M.~D.}\ \bibnamefont
  {Newton}}, \ and\ \bibinfo {author} {\bibfnamefont {D.~V.}\ \bibnamefont
  {Matyushov}},\ }\href@noop {} {\bibfield  {journal} {\bibinfo  {journal} {J.
  Chem. Phys.}\ }\textbf {\bibinfo {volume} {145}},\ \bibinfo {pages} {064504}
  (\bibinfo {year} {2017})}\BibitemShut {NoStop}%
\bibitem [{\citenamefont {Boon}\ and\ \citenamefont {Yip}(1980)}]{Boon:80}%
  \BibitemOpen
  \bibfield  {author} {\bibinfo {author} {\bibfnamefont {J.~P.}\ \bibnamefont
  {Boon}}\ and\ \bibinfo {author} {\bibfnamefont {S.}~\bibnamefont {Yip}},\
  }\href@noop {} {\emph {\bibinfo {title} {Molecular Hydrodynamics}}}\
  (\bibinfo  {publisher} {McGraw-Hill Inc.},\ \bibinfo {year}
  {1980})\BibitemShut {NoStop}%
\bibitem [{\citenamefont {Wilke}, \citenamefont {Chen},\ and\ \citenamefont
  {Bosse}(1999)}]{Wilke:1999iq}%
  \BibitemOpen
  \bibfield  {author} {\bibinfo {author} {\bibfnamefont {S.~D.}\ \bibnamefont
  {Wilke}}, \bibinfo {author} {\bibfnamefont {H.~C.}\ \bibnamefont {Chen}}, \
  and\ \bibinfo {author} {\bibfnamefont {J.}~\bibnamefont {Bosse}},\
  }\href@noop {} {\bibfield  {journal} {\bibinfo  {journal} {Phys. Rev. E}\
  }\textbf {\bibinfo {volume} {60}},\ \bibinfo {pages} {3136} (\bibinfo {year}
  {1999})}\BibitemShut {NoStop}%
\bibitem [{\citenamefont {Munakata}(1975)}]{Munakata:75}%
  \BibitemOpen
  \bibfield  {author} {\bibinfo {author} {\bibfnamefont {T.}~\bibnamefont
  {Munakata}},\ }\href@noop {} {\bibfield  {journal} {\bibinfo  {journal}
  {Prog. Theor. Phys.}\ }\textbf {\bibinfo {volume} {54}},\ \bibinfo {pages}
  {1635} (\bibinfo {year} {1975})}\BibitemShut {NoStop}%
\bibitem [{\citenamefont {Fried}\ and\ \citenamefont
  {Mukamel}(1990)}]{Fried:90}%
  \BibitemOpen
  \bibfield  {author} {\bibinfo {author} {\bibfnamefont {L.~E.}\ \bibnamefont
  {Fried}}\ and\ \bibinfo {author} {\bibfnamefont {S.}~\bibnamefont
  {Mukamel}},\ }\href@noop {} {\bibfield  {journal} {\bibinfo  {journal} {J.\
  Chem.\ Phys.}\ }\textbf {\bibinfo {volume} {93}},\ \bibinfo {pages} {932}
  (\bibinfo {year} {1990})}\BibitemShut {NoStop}%
\bibitem [{\citenamefont {B{{\"o}}ttcher}(1973{\natexlab{b}})}]{Boettcher:73}%
  \BibitemOpen
  \bibfield  {author} {\bibinfo {author} {\bibfnamefont {C.~J.~F.}\
  \bibnamefont {B{{\"o}}ttcher}},\ }\href@noop {} {\emph {\bibinfo {title}
  {Theory of Electric Polarization}}},\ Vol.~\bibinfo {volume} {1}\ (\bibinfo
  {publisher} {Elsevier},\ \bibinfo {address} {Amsterdam},\ \bibinfo {year}
  {1973})\BibitemShut {NoStop}%
\bibitem [{\citenamefont {Madden}\ and\ \citenamefont
  {Kivelson}(1984)}]{Madden:84}%
  \BibitemOpen
  \bibfield  {author} {\bibinfo {author} {\bibfnamefont {P.}~\bibnamefont
  {Madden}}\ and\ \bibinfo {author} {\bibfnamefont {D.}~\bibnamefont
  {Kivelson}},\ }\href@noop {} {\bibfield  {journal} {\bibinfo  {journal} {Adv.
  Chem. Phys.}\ }\textbf {\bibinfo {volume} {56}},\ \bibinfo {pages} {467}
  (\bibinfo {year} {1984})}\BibitemShut {NoStop}%
\bibitem [{\citenamefont {Del~P{\'o}polo}\ and\ \citenamefont
  {Voth}(2004)}]{DelPpolo:2004gx}%
  \BibitemOpen
  \bibfield  {author} {\bibinfo {author} {\bibfnamefont {M.~G.}\ \bibnamefont
  {Del~P{\'o}polo}}\ and\ \bibinfo {author} {\bibfnamefont {G.~A.}\
  \bibnamefont {Voth}},\ }\href@noop {} {\bibfield  {journal} {\bibinfo
  {journal} {J. Phys. Chem. B}\ }\textbf {\bibinfo {volume} {108}},\ \bibinfo
  {pages} {1744} (\bibinfo {year} {2004})}\BibitemShut {NoStop}%
\bibitem [{\citenamefont {Kashyap}\ and\ \citenamefont
  {Biswas}(2010)}]{Kashyap:2010db}%
  \BibitemOpen
  \bibfield  {author} {\bibinfo {author} {\bibfnamefont {H.~K.}\ \bibnamefont
  {Kashyap}}\ and\ \bibinfo {author} {\bibfnamefont {R.}~\bibnamefont
  {Biswas}},\ }\href@noop {} {\bibfield  {journal} {\bibinfo  {journal} {J.
  Phys. Chem. B}\ }\textbf {\bibinfo {volume} {114}},\ \bibinfo {pages} {254}
  (\bibinfo {year} {2010})}\BibitemShut {NoStop}%
\bibitem [{sup()}]{supmatJCP}%
  \BibitemOpen
  \href@noop {} {}\bibinfo {note} {See supplementary material at [{URL} will be
  inserted by AIP] for details of calculations and experimental
  procedures}\BibitemShut {NoStop}%
\bibitem [{\citenamefont {Liu}\ and\ \citenamefont
  {Newton}(1994)}]{Liu:1994ul}%
  \BibitemOpen
  \bibfield  {author} {\bibinfo {author} {\bibfnamefont {Y.-P.}\ \bibnamefont
  {Liu}}\ and\ \bibinfo {author} {\bibfnamefont {M.~D.}\ \bibnamefont
  {Newton}},\ }\href {\doibase 10.1021/j100080a011} {\bibfield  {journal}
  {\bibinfo  {journal} {J. Phys. Chem.}\ }\textbf {\bibinfo {volume} {98}},\
  \bibinfo {pages} {7162} (\bibinfo {year} {1994})}\BibitemShut {NoStop}%
\bibitem [{\citenamefont {Phelps}, \citenamefont {Kornyshev},\ and\
  \citenamefont {Weaver}(1990)}]{Phelps:1990jd}%
  \BibitemOpen
  \bibfield  {author} {\bibinfo {author} {\bibfnamefont {D.~K.}\ \bibnamefont
  {Phelps}}, \bibinfo {author} {\bibfnamefont {A.~A.}\ \bibnamefont
  {Kornyshev}}, \ and\ \bibinfo {author} {\bibfnamefont {M.~J.}\ \bibnamefont
  {Weaver}},\ }\href@noop {} {\bibfield  {journal} {\bibinfo  {journal} {J.
  Phys. Chem.}\ }\textbf {\bibinfo {volume} {94}},\ \bibinfo {pages} {1454}
  (\bibinfo {year} {1990})}\BibitemShut {NoStop}%
\bibitem [{\citenamefont {Krishtalik}, \citenamefont {Alpatova},\ and\
  \citenamefont {Ovsyannikova}(1991)}]{Krishtalik:1991hs}%
  \BibitemOpen
  \bibfield  {author} {\bibinfo {author} {\bibfnamefont {L.~I.}\ \bibnamefont
  {Krishtalik}}, \bibinfo {author} {\bibfnamefont {N.~M.}\ \bibnamefont
  {Alpatova}}, \ and\ \bibinfo {author} {\bibfnamefont {E.~V.}\ \bibnamefont
  {Ovsyannikova}},\ }\href@noop {} {\bibfield  {journal} {\bibinfo  {journal}
  {Electrochim. Acta}\ }\textbf {\bibinfo {volume} {36}},\ \bibinfo {pages}
  {435} (\bibinfo {year} {1991})}\BibitemShut {NoStop}%
\bibitem [{\citenamefont {Baranski}, \citenamefont {Winkler},\ and\
  \citenamefont {Fawcett}(1991)}]{Baranski:1991hy}%
  \BibitemOpen
  \bibfield  {author} {\bibinfo {author} {\bibfnamefont {A.~S.}\ \bibnamefont
  {Baranski}}, \bibinfo {author} {\bibfnamefont {K.}~\bibnamefont {Winkler}}, \
  and\ \bibinfo {author} {\bibfnamefont {W.~R.}\ \bibnamefont {Fawcett}},\
  }\href@noop {} {\bibfield  {journal} {\bibinfo  {journal} {J. Electroanal.
  Chem.}\ }\textbf {\bibinfo {volume} {313}},\ \bibinfo {pages} {367} (\bibinfo
  {year} {1991})}\BibitemShut {NoStop}%
\bibitem [{\citenamefont {Fonseca}\ and\ \citenamefont
  {Ladanyi}(1990)}]{Fonseca:90}%
  \BibitemOpen
  \bibfield  {author} {\bibinfo {author} {\bibfnamefont {T.}~\bibnamefont
  {Fonseca}}\ and\ \bibinfo {author} {\bibfnamefont {B.~M.}\ \bibnamefont
  {Ladanyi}},\ }\href@noop {} {\bibfield  {journal} {\bibinfo  {journal} {J.
  Chem. Phys.}\ }\textbf {\bibinfo {volume} {11}},\ \bibinfo {pages} {8148}
  (\bibinfo {year} {1990})}\BibitemShut {NoStop}%
\bibitem [{\citenamefont {Raineri}, \citenamefont {Resat},\ and\ \citenamefont
  {Friedman}(1992)}]{Raineri:92}%
  \BibitemOpen
  \bibfield  {author} {\bibinfo {author} {\bibfnamefont {F.~O.}\ \bibnamefont
  {Raineri}}, \bibinfo {author} {\bibfnamefont {H.}~\bibnamefont {Resat}}, \
  and\ \bibinfo {author} {\bibfnamefont {H.~L.}\ \bibnamefont {Friedman}},\
  }\href@noop {} {\bibfield  {journal} {\bibinfo  {journal} {J.\ Chem.\ Phys.}\
  }\textbf {\bibinfo {volume} {96}},\ \bibinfo {pages} {3068} (\bibinfo {year}
  {1992})}\BibitemShut {NoStop}%
\bibitem [{\citenamefont {Perng}\ \emph {et~al.}(1996)\citenamefont {Perng},
  \citenamefont {Newton}, \citenamefont {Raineri},\ and\ \citenamefont
  {Friedman}}]{Perng:96}%
  \BibitemOpen
  \bibfield  {author} {\bibinfo {author} {\bibfnamefont {B.-C.}\ \bibnamefont
  {Perng}}, \bibinfo {author} {\bibfnamefont {M.~D.}\ \bibnamefont {Newton}},
  \bibinfo {author} {\bibfnamefont {F.~O.}\ \bibnamefont {Raineri}}, \ and\
  \bibinfo {author} {\bibfnamefont {H.~L.}\ \bibnamefont {Friedman}},\
  }\href@noop {} {\bibfield  {journal} {\bibinfo  {journal} {J. Chem. Phys.}\
  }\textbf {\bibinfo {volume} {104}},\ \bibinfo {pages} {7153} (\bibinfo {year}
  {1996})}\BibitemShut {NoStop}%
\bibitem [{\citenamefont {Reed}, \citenamefont {Madden},\ and\ \citenamefont
  {Papadopoulos}(2008)}]{Reed:2008ez}%
  \BibitemOpen
  \bibfield  {author} {\bibinfo {author} {\bibfnamefont {S.~K.}\ \bibnamefont
  {Reed}}, \bibinfo {author} {\bibfnamefont {P.~A.}\ \bibnamefont {Madden}}, \
  and\ \bibinfo {author} {\bibfnamefont {A.}~\bibnamefont {Papadopoulos}},\
  }\href@noop {} {\bibfield  {journal} {\bibinfo  {journal} {J. Chem. Phys.}\
  }\textbf {\bibinfo {volume} {128}},\ \bibinfo {pages} {124701} (\bibinfo
  {year} {2008})}\BibitemShut {NoStop}%
\bibitem [{\citenamefont {Ben-Amotz}\ and\ \citenamefont
  {Herschbach}(1990)}]{BenAmotz:90}%
  \BibitemOpen
  \bibfield  {author} {\bibinfo {author} {\bibfnamefont {D.}~\bibnamefont
  {Ben-Amotz}}\ and\ \bibinfo {author} {\bibfnamefont {D.~R.}\ \bibnamefont
  {Herschbach}},\ }\href@noop {} {\bibfield  {journal} {\bibinfo  {journal} {J.
  Phys. Chem.}\ }\textbf {\bibinfo {volume} {94}},\ \bibinfo {pages} {1038}
  (\bibinfo {year} {1990})}\BibitemShut {NoStop}%
\bibitem [{\citenamefont {Schmid}\ and\ \citenamefont
  {Matyushov}(1995)}]{DMjpc:95}%
  \BibitemOpen
  \bibfield  {author} {\bibinfo {author} {\bibfnamefont {R.}~\bibnamefont
  {Schmid}}\ and\ \bibinfo {author} {\bibfnamefont {D.~V.}\ \bibnamefont
  {Matyushov}},\ }\href@noop {} {\bibfield  {journal} {\bibinfo  {journal} {J.
  Phys. Chem.}\ }\textbf {\bibinfo {volume} {99}},\ \bibinfo {pages} {2393}
  (\bibinfo {year} {1995})}\BibitemShut {NoStop}%
\bibitem [{\citenamefont {Mezger}\ \emph {et~al.}(2008)\citenamefont {Mezger},
  \citenamefont {Schroder}, \citenamefont {Reichert}, \citenamefont {Schramm},
  \citenamefont {Okasinski}, \citenamefont {Schoder}, \citenamefont
  {Honkimaki}, \citenamefont {Deutsch}, \citenamefont {Ocko}, \citenamefont
  {Ralston}, \citenamefont {Rohwerder}, \citenamefont {Stratmann},\ and\
  \citenamefont {Dosch}}]{Mezger:08}%
  \BibitemOpen
  \bibfield  {author} {\bibinfo {author} {\bibfnamefont {M.}~\bibnamefont
  {Mezger}}, \bibinfo {author} {\bibfnamefont {H.}~\bibnamefont {Schroder}},
  \bibinfo {author} {\bibfnamefont {H.}~\bibnamefont {Reichert}}, \bibinfo
  {author} {\bibfnamefont {S.}~\bibnamefont {Schramm}}, \bibinfo {author}
  {\bibfnamefont {J.~S.}\ \bibnamefont {Okasinski}}, \bibinfo {author}
  {\bibfnamefont {S.}~\bibnamefont {Schoder}}, \bibinfo {author} {\bibfnamefont
  {V.}~\bibnamefont {Honkimaki}}, \bibinfo {author} {\bibfnamefont
  {M.}~\bibnamefont {Deutsch}}, \bibinfo {author} {\bibfnamefont {B.~M.}\
  \bibnamefont {Ocko}}, \bibinfo {author} {\bibfnamefont {J.}~\bibnamefont
  {Ralston}}, \bibinfo {author} {\bibfnamefont {M.}~\bibnamefont {Rohwerder}},
  \bibinfo {author} {\bibfnamefont {M.}~\bibnamefont {Stratmann}}, \ and\
  \bibinfo {author} {\bibfnamefont {H.}~\bibnamefont {Dosch}},\ }\href@noop {}
  {\bibfield  {journal} {\bibinfo  {journal} {Science}\ }\textbf {\bibinfo
  {volume} {322}},\ \bibinfo {pages} {424} (\bibinfo {year}
  {2008})}\BibitemShut {NoStop}%
\bibitem [{\citenamefont {Sikes}(2001)}]{Sikes:2001hp}%
  \BibitemOpen
  \bibfield  {author} {\bibinfo {author} {\bibfnamefont {H.~D.}\ \bibnamefont
  {Sikes}},\ }\href@noop {} {\bibfield  {journal} {\bibinfo  {journal}
  {Science}\ }\textbf {\bibinfo {volume} {291}},\ \bibinfo {pages} {1519}
  (\bibinfo {year} {2001})}\BibitemShut {NoStop}%
\bibitem [{\citenamefont {Raineri}\ and\ \citenamefont
  {Friedman}(1999)}]{Raineri:99}%
  \BibitemOpen
  \bibfield  {author} {\bibinfo {author} {\bibfnamefont {F.~O.}\ \bibnamefont
  {Raineri}}\ and\ \bibinfo {author} {\bibfnamefont {H.~L.}\ \bibnamefont
  {Friedman}},\ }\href@noop {} {\bibfield  {journal} {\bibinfo  {journal} {Adv.
  Chem. Phys.}\ }\textbf {\bibinfo {volume} {107}},\ \bibinfo {pages} {81}
  (\bibinfo {year} {1999})}\BibitemShut {NoStop}%
\bibitem [{\citenamefont {Vatamanu}, \citenamefont {Borodin},\ and\
  \citenamefont {Smith}(2010)}]{Vatamanu:2010we}%
  \BibitemOpen
  \bibfield  {author} {\bibinfo {author} {\bibfnamefont {J.}~\bibnamefont
  {Vatamanu}}, \bibinfo {author} {\bibfnamefont {O.}~\bibnamefont {Borodin}}, \
  and\ \bibinfo {author} {\bibfnamefont {G.~D.}\ \bibnamefont {Smith}},\
  }\href@noop {} {\bibfield  {journal} {\bibinfo  {journal} {Phys. Chem. Chem.
  Phys.}\ }\textbf {\bibinfo {volume} {12}},\ \bibinfo {pages} {170} (\bibinfo
  {year} {2010})}\BibitemShut {NoStop}%
\bibitem [{\citenamefont {Rovere}\ and\ \citenamefont
  {Tosi}(1986)}]{Rovere:1986iq}%
  \BibitemOpen
  \bibfield  {author} {\bibinfo {author} {\bibfnamefont {M.}~\bibnamefont
  {Rovere}}\ and\ \bibinfo {author} {\bibfnamefont {M.~P.}\ \bibnamefont
  {Tosi}},\ }\href@noop {} {\bibfield  {journal} {\bibinfo  {journal} {Rep.
  Prog. Phys.}\ }\textbf {\bibinfo {volume} {49}},\ \bibinfo {pages} {1001}
  (\bibinfo {year} {1986})}\BibitemShut {NoStop}%
\bibitem [{\citenamefont {Bazant}, \citenamefont {Storey},\ and\ \citenamefont
  {Kornyshev}(2011)}]{Bazant:2011ha}%
  \BibitemOpen
  \bibfield  {author} {\bibinfo {author} {\bibfnamefont {M.~Z.}\ \bibnamefont
  {Bazant}}, \bibinfo {author} {\bibfnamefont {B.~D.}\ \bibnamefont {Storey}},
  \ and\ \bibinfo {author} {\bibfnamefont {A.~A.}\ \bibnamefont {Kornyshev}},\
  }\href@noop {} {\bibfield  {journal} {\bibinfo  {journal} {Phys. Rev. Lett.}\
  }\textbf {\bibinfo {volume} {106}},\ \bibinfo {pages} {1247} (\bibinfo {year}
  {2011})}\BibitemShut {NoStop}%
\bibitem [{\citenamefont {Fedorov}\ and\ \citenamefont
  {Kornyshev}(2014)}]{Fedorov:2014er}%
  \BibitemOpen
  \bibfield  {author} {\bibinfo {author} {\bibfnamefont {M.~V.}\ \bibnamefont
  {Fedorov}}\ and\ \bibinfo {author} {\bibfnamefont {A.~A.}\ \bibnamefont
  {Kornyshev}},\ }\href@noop {} {\bibfield  {journal} {\bibinfo  {journal}
  {Chem. Rev.}\ }\textbf {\bibinfo {volume} {114}},\ \bibinfo {pages} {2978}
  (\bibinfo {year} {2014})}\BibitemShut {NoStop}%
\bibitem [{\citenamefont {Nikitina}\ \emph
  {et~al.}(2014{\natexlab{b}})\citenamefont {Nikitina}, \citenamefont
  {Kislenko}, \citenamefont {Nazmutdinov}, \citenamefont {Bronshtein},\ and\
  \citenamefont {Tsirlina}}]{Nikitina:2014ei}%
  \BibitemOpen
  \bibfield  {author} {\bibinfo {author} {\bibfnamefont {V.~A.}\ \bibnamefont
  {Nikitina}}, \bibinfo {author} {\bibfnamefont {S.~A.}\ \bibnamefont
  {Kislenko}}, \bibinfo {author} {\bibfnamefont {R.~R.}\ \bibnamefont
  {Nazmutdinov}}, \bibinfo {author} {\bibfnamefont {M.~D.}\ \bibnamefont
  {Bronshtein}}, \ and\ \bibinfo {author} {\bibfnamefont {G.~A.}\ \bibnamefont
  {Tsirlina}},\ }\href@noop {} {\bibfield  {journal} {\bibinfo  {journal} {J.
  Phys. Chem. C}\ }\textbf {\bibinfo {volume} {118}},\ \bibinfo {pages} {6151}
  (\bibinfo {year} {2014}{\natexlab{b}})}\BibitemShut {NoStop}%
\bibitem [{com()}]{comLambda}%
  \BibitemOpen
  \href@noop {} {}\bibinfo {note} {The reorganization energies reported in
  Ref.\ \onlinecite{Nikitina:2014ei} are calculated by dividing the values from
  MD simulations by $\epsilon_\infty$.}\BibitemShut {Stop}%
\bibitem [{\citenamefont {Nakanishi}\ and\ \citenamefont
  {Sokolov}(2014)}]{Nakanishi:2014gy}%
  \BibitemOpen
  \bibfield  {author} {\bibinfo {author} {\bibfnamefont {M.}~\bibnamefont
  {Nakanishi}}\ and\ \bibinfo {author} {\bibfnamefont {A.~P.}\ \bibnamefont
  {Sokolov}},\ }\href@noop {} {\bibfield  {journal} {\bibinfo  {journal} {J.
  Non-Cryst. Solids}\ }\textbf {\bibinfo {volume} {407}},\ \bibinfo {pages}
  {478} (\bibinfo {year} {2014})}\BibitemShut {NoStop}%
\bibitem [{\citenamefont {Palmer}(1982)}]{Palmer:82}%
  \BibitemOpen
  \bibfield  {author} {\bibinfo {author} {\bibfnamefont {R.~G.}\ \bibnamefont
  {Palmer}},\ }\href@noop {} {\bibfield  {journal} {\bibinfo  {journal} {Adv.
  Phys.}\ }\textbf {\bibinfo {volume} {31}},\ \bibinfo {pages} {669} (\bibinfo
  {year} {1982})}\BibitemShut {NoStop}%
\bibitem [{\citenamefont {Crisanti}\ and\ \citenamefont
  {Ritort}(2000)}]{Crisanti:00}%
  \BibitemOpen
  \bibfield  {author} {\bibinfo {author} {\bibfnamefont {A.}~\bibnamefont
  {Crisanti}}\ and\ \bibinfo {author} {\bibfnamefont {F.}~\bibnamefont
  {Ritort}},\ }\href@noop {} {\bibfield  {journal} {\bibinfo  {journal}
  {Physica A}\ }\textbf {\bibinfo {volume} {280}},\ \bibinfo {pages} {155}
  (\bibinfo {year} {2000})}\BibitemShut {NoStop}%
\end{thebibliography}

%

\end{document}